\documentclass[sigconf]{acmart}
\pdfoutput=1

\usepackage{xspace}
\usepackage{color}
\usepackage{url}
\usepackage{subcaption}
\usepackage{paralist}
\usepackage{wrapfig}
\usepackage{multirow}
\usepackage{listings}
\usepackage[ruled,vlined]{algorithm2e}
\usepackage{booktabs}
\usepackage{makecell}
\usepackage{boxedminipage}

\newcommand{\bheading}[1]{{\vspace{4pt}\noindent{\textbf{#1}}}}

\newcounter{note}[section]

\newcommand{\secref}[1]{\mbox{Section~\ref{#1}}\xspace}

\newcommand{\figref}[1]{\mbox{Figure~\ref{#1}}}

\newcommand{\tabref}[1]{\mbox{Table~\ref{#1}}}
\newcommand{\appref}[1]{\mbox{Appendix~\ref{#1}}}
\newcommand{\ignore}[1]{}

\newcommand{\etc}{\emph{etc}\xspace}
\newcommand{\ie}{\emph{i.e.}\xspace}
\newcommand{\eg}{\emph{e.g.}\xspace}

\newcommand{\etal}{\emph{et al.}\xspace}

\usepackage{pifont}
\newcommand{\cmark}{\ding{51}}
\newcommand{\xmark}{\ding{55}}

\newcounter{packednmbr}

\newenvironment{packeditemize}{
\begin{list}{$\bullet$}{
\setlength{\labelwidth}{8pt}
\setlength{\itemsep}{0pt}
\setlength{\leftmargin}{\labelwidth}
\addtolength{\leftmargin}{\labelsep}
\setlength{\parindent}{0pt}
\setlength{\listparindent}{\parindent}
\setlength{\parsep}{2pt}
\setlength{\topsep}{3pt}}}{\end{list}}

\definecolor{dkgreen}{rgb}{0,0.6,0}
\definecolor{gray}{rgb}{0.5,0.5,0.5}
\definecolor{mauve}{rgb}{0.58,0,0.82}

\newcommand{\flushreload}{\textsc{Flush-Reload}\xspace}

\newcommand{\vulNames}{sensitive control-flow vulnerabilities\xspace}
\newcommand{\VulNames}{Sensitive Control-Flow Vulnerabilities\xspace}
\newcommand{\attackNames}{control-flow inference attacks\xspace}
\newcommand{\AttackNames}{Control-Flow Inference Attacks\xspace}
\newcommand{\frameName}{\textsc{Stacco}\xspace}

\begin{document}

\copyrightyear{2017}
\acmYear{2017}
\setcopyright{rightsretained}
\acmConference{CCS '17}{October 30-November 3, 2017}{Dallas, TX, USA}
\acmDOI{10.1145/3133956.3134016}
\acmISBN{978-1-4503-4946-8/17/10}

%\title{Oracle Attacks against SSL/TLS Implementations in Secure Enclaves}
\title{\frameName: Differentially Analyzing Side-Channel Traces for Detecting SSL/TLS
Vulnerabilities in Secure Enclaves}

\author{Yuan Xiao}
\affiliation{The Ohio State University}
\email{xiao.465@osu.edu}

\author{Mengyuan Li}
\affiliation{The Ohio State University}
\email{li.7533@osu.edu}

\author{Sanchuan Chen}
\affiliation{The Ohio State University}
\email{chen.4825@osu.edu}

\author{Yinqian Zhang}
\affiliation{The Ohio State University}
\email{yinqian@cse.ohio-state.edu}

\begin{abstract}
Intel Software Guard Extension (SGX) offers software applications a shielded
execution environment, dubbed \textit{enclave}, to protect their confidentiality
and integrity from malicious operating systems. As processors with this extended
feature become commercially available, many new software applications are
developed to enrich to the SGX-enabled software ecosystem. One important primitive for
these applications is a secure communication channel between the enclave and a
remote trusted party.  The SSL/TLS protocol, which is the \textit{de facto}
standard for protecting transport-layer network communications, has been broadly
regarded a natural choice for such purposes.  However, in this paper, we show
that the marriage between SGX and SSL may not be a smooth sailing. 

Particularly, we consider a category of side-channel attacks against SSL/TLS
implementations in secure enclaves, which we call the control-flow inference
attacks. In these attacks, the malicious operating system kernel may perform a
powerful \textit{man-in-the-kernel} attack to collect execution traces of the
enclave programs at the page level, the cacheline level, or the branch level,
while positioning itself in the middle of the two communicating parties. At the
center of our work is a \textit{differential analysis framework}, dubbed
\frameName, to dynamically analyze the SSL/TLS implementations and detect
vulnerabilities---discernible execution traces---that can be exploited as
decryption oracles.  Surprisingly, in spite of the prevailing constant-time
programming paradigm adopted by many cryptographic libraries, we found exploitable
vulnerabilities in the latest versions of all the SSL/TLS libraries we have
examined. 

To validate the detected vulnerabilities, we developed a man-in-the-kernel
adversary to demonstrate Bleichenbacher attacks against the latest OpenSSL
library running in the SGX enclave (with the help of Graphene) and completely
broke the \texttt{PreMasterSecret} encrypted by a 4096-bit RSA public key with
only 57,286 queries. We also conducted CBC padding oracle attacks against the
latest GnuTLS running in Graphene-SGX and an open-source SGX-implementation of
mbedTLS (\ie, mbedTLS-SGX) that runs directly inside the enclave, and showed
that it only needs 48,388 and 25,717 queries, respectively, to break one block of
AES ciphertext.  Empirical evaluation suggests these man-in-the-kernel attacks
can be completed within one or two hours.  

%Our results reveal the insufficient understanding of side-channel security in
%SGX settings.  and our study will provoke discussions on the secure
%implementation and adoption of SSL/TLS in secure enclaves.  

\end{abstract}

\maketitle

%section introduction
\section{Introduction}
\label{sec:intro} %comments

Software applications' security heavily depends on the security of the
underlying system software. In traditional computing environments, if the
operating system is compromised, the security of the applications it supports is
also compromised. Therefore, the trusted computing base (TCB) of software
applications include not only the software itself but also the underlying
system software and hardware. 

To reduce the TCB of some applications that contain sensitive code and data,
academic researchers have proposed many software systems to support
\textit{shielded execution}---\ie, execution of a piece of code whose
confidentiality and integrity is protected from an untrusted system software
(\eg,~\cite{Lie:2003:IUO, TaMin:2006:SIM, Chen:2007:tamper,
Chen:2008:overshadow, Mccune:2008:flicker, Yang:2008:perpage, Ports:2008:TAS,
Zhang:2011:cloudvisor, Hofmann:2013:inktag, Cheng:2013:appshield,
Criswell:2014:virtualghost, Li:2014:twoway}). Most of these systems adopted a
hypervisor-based approach to protecting the memory of victim applications
against attacks from malicious operating systems.  Although promising, these
academic prototypes have yet to see the light of real-world adoption.  Not until
the advent of Intel Software Guard eXtension (SGX)~\cite{IntelSGXManual}, a
hardware extension available in the most recent Intel processors, did the
concept of shielded execution become practical to real-world applications. SGX
enforces both confidentiality and integrity of userspace programs by isolating
regions of their memory space (\ie, \textit{enclaves}) from other software
components, including the most privileged system software--no memory reads or
writes can be performed inside the enclaves by any external software, regardless
of its privilege level.  As such, SGX greatly reduces the TCB of the shielded
execution, enabling a wide range of
applications~\cite{Hoekstra:2013:sgxsolution, Baumann:2015:haven,
Schuster:2015:vc3, Zhang:2016:towncrier, Ohrimenko:2016:OMP, Sinha:2016:DVM}.

In typical application scenarios~\cite{Hoekstra:2013:sgxsolution,
Baumann:2015:haven, Zhang:2016:towncrier}, shielded execution does not work completely 
alone; it communicates with remote trusted parties using secure channels,
\eg, SSL/TLS protocols. Secure Sockets Layer (SSL) and its successor, Transport
Layer Security (TLS), are transport-layer security protocols that provide secure
communication channels using a set of cryptographic primitives. SSL/TLS
protocols are expected, as part of their design goals, to prevent
man-in-the-middle attackers who are capable of eavesdropping, intercepting,
replaying, modifying and injecting network packets between the two communicating
parties. Therefore, applications of Intel SGX~\cite{Hoekstra:2013:sgxsolution,
Baumann:2015:haven, Zhang:2016:towncrier} typically regard SSL/TLS modules inside
SGX enclaves as basic security primitives to establish
end-to-end communication security.

Attacks against the SSL/TLS protocol have been reported over the years,
unfortunately. One important category of these attacks is oracle
attacks~\cite{Clark:2013:SSH}.  In an oracle attack, the adversary interactively
and adaptively queries a vulnerable SSL/TLS implementation and uses the
response (or some side-channel information, \eg, the latency of the response) as an oracle to
break the encryption.  Well-known examples of oracle attacks include the Lucky
Thirteen Attack~\cite{Fardan:2013:lucky}, the Bleichenbacher
attack~\cite{bleichenbacher1998chosen}, the DROWN attack~\cite{aviramdrown},
the POODLE attack~\cite{Moller:2014:poodle}, \etc. Prior demonstration of these
attacks have shown that they enable network attackers to decrypt arbitrary
messages of the SSL record protocol or decrypt the \textit{PreMasterSecret} of
the SSL handshake protocol. We will detail these attacks in
\secref{sec:background}.  Due to the broad adoption of the SSL/TLS protocol (\eg,
in HTTPS, secure email exchanges), any of these attacks is devastating and
easily headlines of the security news (\eg,~\cite{bbc:drown}).  Accordingly, the
SSL/TLS protocol and its implementations have been frequently updated after the
publicity of these attacks. A commonly used solution is to \textit{hide} the
oracles. For example, in cases where the oracle is the SSL Alert message
indicating padding errors, the error message can be unified to conceal the real
reason for the errors~\cite{rescorla2006transport, dierks2008transport} (so that the
adversary cannot differentiate padding errors and MAC errors, see
\secref{sec:background}). As of today, almost all widely used SSL/TLS
implementations are resilient to oracle attacks because the oracles have been
successfully hidden from the network attackers~\cite{rescorla2006transport,
dierks2008transport, opensslpatch, gnutlssecurity}.

However, adoption of SSL/TLS in SGX enclaves brings new security challenges.
Although SGX offers confidentiality protection, through memory isolation and
encryption, to code and data inside secure enclaves, it has been shown 
vulnerable to side-channel attacks~\cite{Xu:2015:controlled, Shinde:2015:PYF,
Lee:2016:SGXbranch}. Side-channel attacks are a type of security attacks against
the confidentiality of a system or application by making inferences from
measurements of observable side-channel events. These attacks have been studied
in the past twenty years in multiple contexts, most noticeably in desktop
computers, cloud servers, and mobile devices 
%from one unprivileged software against another unprivileged software or system
%software by observing interference via 
where CPU micro-architectures~\cite{Zhang:2014:CSA, zhang2016return}, software
data structures~\cite{Rane:2015:raccoon, jana2012memento}, or other system
resources are shared between mutually-distrusting software components. What
makes side-channel attacks on SGX different is that these attacks can be
performed by the privileged system software, which enables many new attack
vectors. For example, Xu \etal~\cite{Xu:2015:controlled}
demonstrated that by manipulating page table entries of the memory pages of
secure enclaves, an adversary with system privilege could enforce page faults
during the execution of enclave programs, thus collecting traces of memory
accesses at the page-granularity. Recently, Lee \etal~\cite{Lee:2016:SGXbranch}
demonstrated that the control flow of enclave programs can be precisely traced at
every branch instruction by exploiting the shared Branch Prediction Units (BPU). 

The \textit{key insight} of this paper is that while SSL/TLS is designed to
defend against \textit{man-in-the-middle} attacks, its implementation in SGX
enclaves must tackle a stronger \textit{man-in-the-kernel} adversary who is
capable of not only positioning himself in the \textit{middle} of the two
communicating parties, but controlling the underlying operating system kernel and
manipulating system resources to collect execution traces of the enclave
programs from various side channels.  Particularly, we show that the powerful
man-in-the-kernel attackers can create new decryption oracles from the
state-of-the-art SSL/TLS implementations and resurrect the Bleichenbacher attack
and CBC padding oracle attacks against SGX enclaves.

%Different from previously demonstrated attacks on SGX that only show successful
%attacks against old, vulnerable versions of OpenSSL (\eg, vulnerable RSA
%implementation in version xx.xx.xx~\cite{}; vulnerable ECDSA implementation in
%version xx.xx.xx~\cite{}), and mbedTLS (), and against libraries that are
%unlikely representative of typical use cases of SGX (\eg, input to the FreeType
%font rendering engine~\cite{Xu:2015:controlled}, the Hunspell spell checker~
%\cite{Xu:2015:controlled}, LIBSVM
%classifier~\cite{zhang2012cross},)

\bheading{\frameName.}
At the core of our work is the Side-channel Trace Analyzer for finding
Chosen-Ciphertext Oracles (\frameName), which is a software framework for
conducting differential analysis on the SSL/TLS implementations to detect
\textit{\vulNames} that can be exploited to create decryption oracles for CBC
padding oracle attacks and Bleichenbacher attacks.  Particularly, to enable
automated large-scale analysis of various off-the-shelf SSL/TLS libraries, we
built \frameName on top of a dynamic instrumentation engine (\ie,
Pin~\cite{luk2005pin}) and an open-source SSL/TLS packet generation tool (\ie,
TLS-Attacker~\cite{Somorovsky:2016:tlsattacker}), so that we can perform
standard tests to multiple libraries in an automated manner. To understand the
exploitability of the vulnerabilities, we also modeled three types of
control-flow inference attacks, including page-level
attacks~\cite{Xu:2015:controlled, Shinde:2015:PYF}, cacheline-level
attacks~\cite{Schwarz:2017:MGE, Brasser:2017:SGE} and branch-level
attacks~\cite{Lee:2016:SGXbranch}, and empowered \frameName to analyze
vulnerabilities on each of these levels.  Our analysis results suggest all the
popular open-source SSL/TLS libraries we have examined are vulnerable to both
types of oracle attacks, raising the questions of secure development and
deployment of SSL/TLS protocols inside SGX enclaves.

To validate the vulnerabilities identified by \frameName, we demonstrated
several such \textit{man-in-the-kernel} attacks against the latest versions of
popular cryptographic libraries: Particularly, we implemented a Bleichenbarcher
attack against the latest OpenSSL library~\cite{openssl102} running in the SGX
enclaves (with the help of Graphene-SGX~\cite{tsai2014cooperation}, a library OS
that supports unmodified applications to run inside SGX enclaves) and completely
broke the \texttt{PreMasterSecret} encrypted by a 4096-bit RSA public key with
only 57,286 queries. We also conducted CBC padding oracle attacks against the
latest GnuTLS~\cite{gnutlsnews} running in Graphene-SGX and an open-source
SGX-implementation of mbedTLS~\cite{mbedtlssgx} that runs directly inside the
enclave, and showed that it only needs 48,388 and 25,717 queries, respectively, to
break one block of AES ciphertext from TLS connections using these libraries.
Empirical evaluation suggests these man-in-the-kernel attacks can be completed
within one or two hours. These demonstrated attacks not only provide evidence that
\frameName can effectively identify exploitable \vulNames in SSL/TLS
implementations, but also suggest these oracle attacks conducted in a
man-in-the-kernel manner are efficient for practical security intrusion.

\bheading{Responsible disclosure.} We have reported the vulnerabilities and demonstrated oracle
attacks to Intel, OpenSSL, GnuTLS, mbedTLS.

\bheading{Contributions of this work include:}

\begin{packeditemize}

\item The first study of critical side-channel threats against SSL/TLS
implementations in SGX enclaves that lead to complete compromises of
SSL/TLS-protected secure communications.

\item The design and implementation of \frameName, a differential analysis
framework for detecting \vulNames in SSL/TLS implementations, which also
entails: 

\item A systematic characterization of \attackNames against SGX enclaves (\eg,
page-level attacks, the cacheline-level attacks, and branch-level attacks),
which empowers \frameName to analyze the vulnerability with abstracted attacker
models.  

\item A measurement study of the latest versions of popular SSL/TLS libraries
using \frameName that shows that all of them, including OpenSSL, GnuTLS,
mbedTLS, WolfSSL, and LibreSSL, are vulnerable to \attackNames and exploitable
in oracle attacks.

\item An empirical \textit{man-in-the-kernel} demonstration of oracle attacks
against the latest version of OpenSSL and GnuTLS running inside
Graphene-SGX and an open-source SGX-implementation of mbedTLS, showing that such
attacks are highly efficient on \textit{real} SGX hardware.

\end{packeditemize}

\bheading{Roadmap.} The rest of this paper is outlined as follows.
\secref{sec:background} introduces related background concepts.
\secref{sec:threat} systematically characterizes \attackNames.
\secref{sec:detect} describes a differential analysis framework for detecting
\vulNames in SSL/TLS implementations. We demonstrate oracle attacks against some
of the vulnerable SSL/TLS implementations to validate these detected
vulnerabilities in \secref{sec:attack}, and then discuss countermeasures in
\secref{sec:discuss}. In \secref{sec:related}, we briefly summarize related work
in the field.  \secref{sec:conclude} concludes our paper.

\section{Background}
\label{sec:background}

\subsection{Intel Software Guard Extension}
Intel SGX is a new processor architecture extension that is available on the most
recent Intel processors (\eg, Skylake and later processor families). It aims to
protect the confidentiality and integrity of code and data of sensitive
applications against malicious system software~\cite{IntelSGXManual}.  The
protection is achieved through a set of security primitives, such as memory
isolation and encryption, sealed storage, remote attestation, \etc. In this
section, we briefly introduce some of the key features of Intel SGX that is
relevant to this paper. 

\bheading{Memory isolation and encryption.}
Intel SGX reserves a range of continuous physical memory exclusively for
enclaves. This memory range is called Enclave Page Cache (EPC), which is a
subset of Processor Reserved Memory (PRM).  The EPC is managed in similar ways
to regular physical memory and is divided into 4KB pages. Correspondingly,  a
range of virtual addresses, called Enclave Linear Address Range (ELRANGE), is
reserved in the virtual address space of the applications. The page tables
responsible for address translation are managed by the untrusted operating
system. Therefore, the mapping of each virtual memory page to the physical
memory, access permissions, cacheability, \etc., can be controlled by the system
software that is potentially malicious. To maintain the integrity of the page
tables, the memory access permission dictated by the developers are recorded,
upon enclave initiation, in Enclave Page Cache Map (EPCM), which is also part of
PRM (thus protected from the malicious system software). During the address
translation, EPCM is consulted to enforce access permission by bitwise-AND the
set of permissions in the EPCM entries and those in the page table entries. 

The memory management unit (MMU) enforces integrity and confidentiality of EPC
pages. Only code running in enclave mode can access virtual memory pages in the
ELRANGE that are mapped to the EPC. Each EPC page has at most one owner at a
time, and the EPCM serves as a revert page table that records virtual address
space of each enclave that maps to the corresponding EPC page. An EPC page can
be evicted and stored in the regular physical memory region. Evicted EPC pages are
encrypted by Memory Encryption Engine (MEE) to guarantee their confidentiality.  

%\bheading{Side-channel attacks against SGX.} Despite the enhanced memory
%isolation and encryption in SGX, protecting the confidentiality of data and code
%inside the secure enclave is not easy. It has been demonstrated in previous
%studies that a variety of side-channel attacks can be performed to infer
%sensitive information from enclaves during their
%execution~\cite{Xu:2015:controlled, Shinde:2015:PYF, Lee:2016:SGXbranch}. This
%paper systematically studies these side-channel attacks and discuss their
%implications to secure implementation of SSL/TLS protocols inside SGX enclaves.   

\subsection{SSL/TLS}

\begin{wrapfigure}{r}{0.45\columnwidth}
    \centering
        \includegraphics[width=0.44\columnwidth]{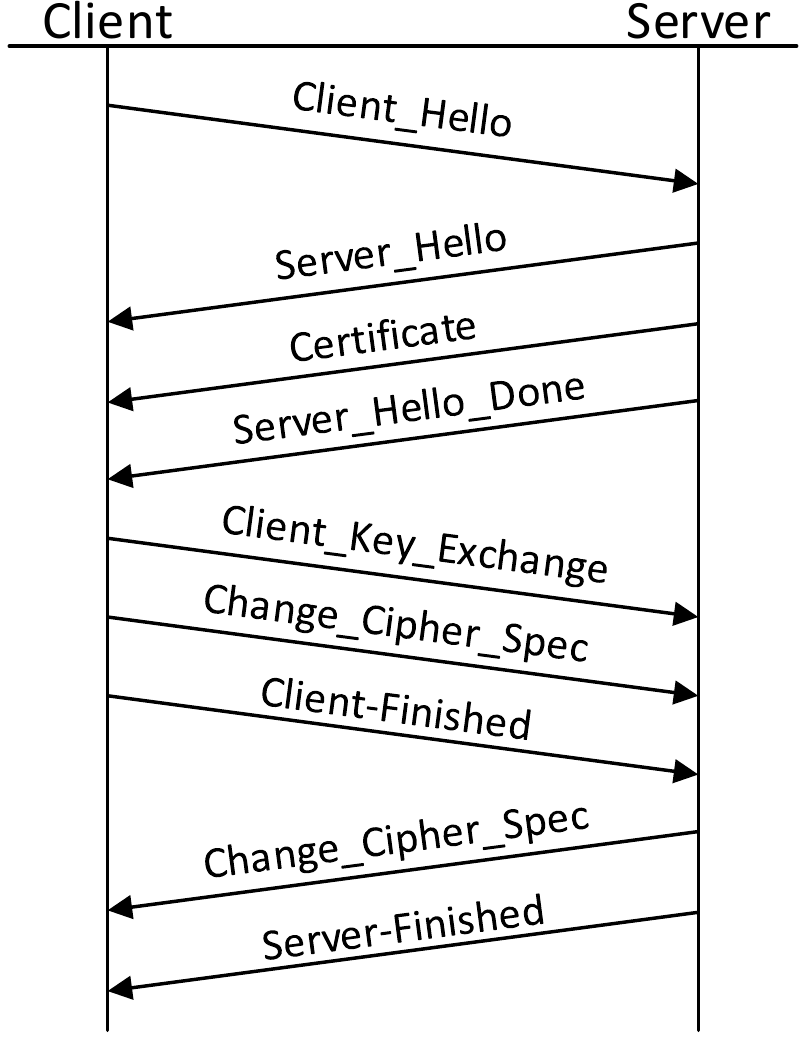}
    \caption{The SSL handshake protocol.}
    \label{fig:handshake}
\end{wrapfigure}

Secure Sockets Layer (SSL) is a general purpose security protocol proposed by
Netscape Communications in 1994, which was designed to transparently protect the
confidentiality and integrity of the network communications between applications
running on top of the TCP layer. Due to security flaws, SSL v1.0 was never
released to the public. SSL v2.0 was released in 1995 and deprecated in 2011; SSL
v3.0 was released in 1996 and deprecated in 2015 (after the publicity of POODLE
attacks~\cite{Moller:2014:poodle}). Its successor protocols, Transport Layer
Security (TLS) v1.0, v1.1, and v1.2, were released in 1999, 2006 and 2008,
respectively. They are all still being used broadly. TLS v1.3 is still a working
draft as of August 2017. SSL and TLS are referred together as the SSL/TLS protocol.
SSL/TLS has two sub-protocols, the \textit{handshake protocol}, and the
\textit{record layer protocol}. The handshake protocol negotiates security
primitives (ciphers, their parameters, and cryptographic keys), and the record
layer protocol uses the negotiated security primitives for encryption and
authentication of the payload data, such as HTTP, IMAP, SMTP, POP3, \etc.

\bheading{Handshake protocol.}
The SSL handshake protocol allows the communicating server and client to
authenticate each other and negotiate an algorithm for message encryption and
integrity protection. The protocol is illustrated in \figref{fig:handshake}.
The client initiates the SSL connection with a \texttt{ClientHello} message,
which tells the server the maximum SSL version it supports, a 28-byte random
value, the identifier of the SSL session that this current SSL connection is
associated to, the set of supported ciphers, and the compression algorithms. The
server, upon receiving the client's request, responds with a \texttt{ServerHello}
message, with the same set of information from the server. The server will then
send a \texttt{Certificate} message, if it is the first connection of the
session, to offer its certificate to the client. If the certificate used by the
server is a certificate that uses Digital Signature Algorithms (DSA) or a
signing-only RSA certificate, it does not have a key that can be used for
encryption purposes. In this case, the server will send a
\texttt{ServerKeyExchange} message to inform the client its Diffie-Hellman (DH)
parameters to perform the key exchange.

Upon receiving the \texttt{ServerHelloDone} message from the server, the client
sends a \texttt{ClientKeyExchange} message to the server. If RSA key exchanges
are used, the \texttt{PreMasterSecret} will be encrypted using the RSA public
key embedded in the certificate and sent along with the message; if
Diffie-Hellman key exchange algorithms are used, this message will only include
the client's DH parameters---the \texttt{PreMasterSecret} is calculated by the
server and client respectively without being sent over the network. After this
step, the server and the client already share the secrets for generating the
symmetric encryption keys and Message Authentication Code (MAC) keys. The
\texttt{ChangeCipherSpec} messages from the client and the server notify the
other party about the forthcoming changes to the cipher algorithms that have
just been negotiated.

\begin{figure}
    \centering
	\includegraphics[width=0.85\columnwidth]{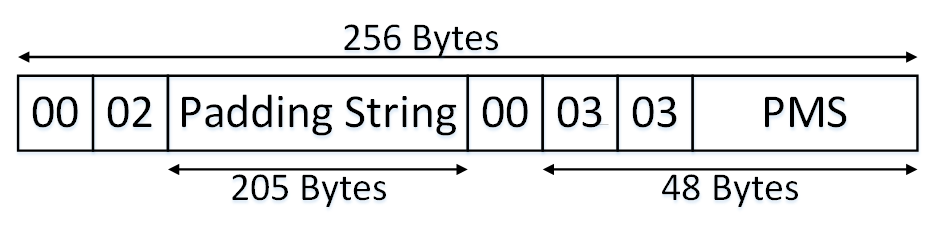}
    \caption{Format of the plaintext of the \texttt{ClientKeyExchange} message
(with 2048-bit RSA keys).}
    \label{fig:blockformat}
\end{figure}

Particularly, when RSA-based key exchange method is selected, the
\texttt{PreMasterSecret} is encrypted using the server's public RSA key. The
format of the plaintext message of \texttt{ClientKeyExchange} conforms to a
variant of PKCS\#1 v1.5 format (shown in \figref{fig:blockformat}): it must
start with $0x0002$ which is followed by 205 bytes of non-zero paddings provided
that the total message is 256-byte long (determined by the size of the server's
private key). Then a $0x00$ byte following the padding is regarded as the
segmentation mark, and the 48-byte \texttt{PreMasterSecret} is attached at the
end. According to RFC5246 (TLS v1.2), in order to defeat Bleichenbacher Attacks,
which we will detail shortly, the server first generates a random value, and
then decrypts the \texttt{ClientKeyExchange} message. If the decrypted data does
not conform to the PKCS\#1 standard or the length of the \texttt{PreMasterSecret} is
incorrect, the random value will be used for the rest of the computation, as if
the decryption was successful.

TLS v1.0, v1.1, and v1.2 support a variety of cipher suites. For example,
\texttt{TLS\_RSA\_WITH\_AES\_128\_CBC\_SHA} is one of the cipher suites which employs
RSA for both authentication and key exchange, the symmetric encryption uses the AES
block cipher in Cipher Block Chaining (CBC) mode, and SHA-1 based
HMAC is used for integrity protection of the payload. Other key exchange
algorithms can also be specified. For instance, 
\texttt{TLS\_ECDHE\_ECDSA} uses elliptic curve
Diffie-Hellman key exchange and Elliptic Curve Digital Signature Algorithm for
authentication.

\begin{figure}[tbh]
\centering
\includegraphics[width=0.95\columnwidth]{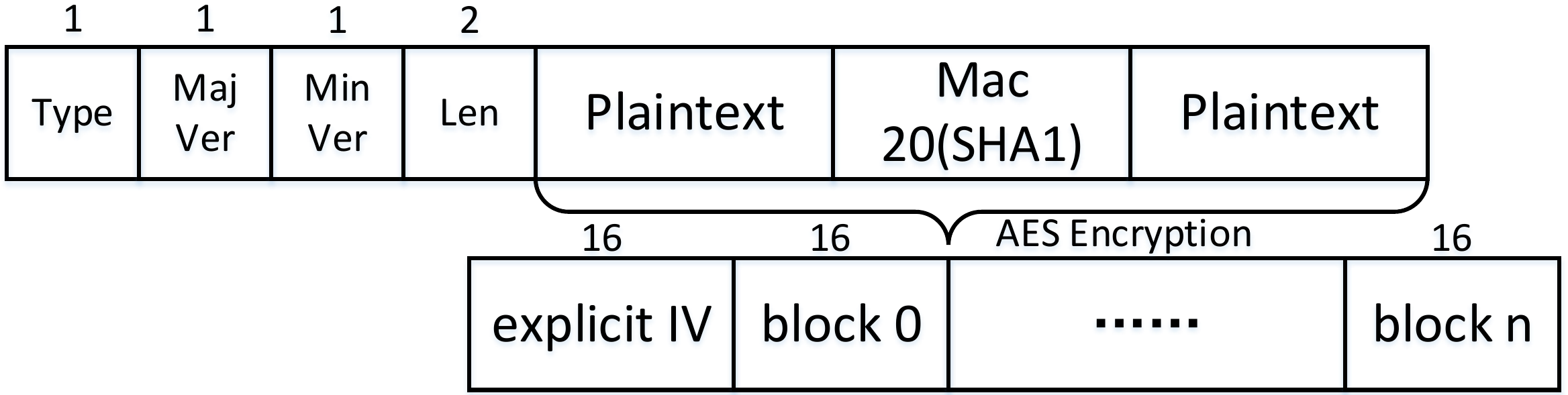}
\caption{Packet format of SSL/TLS payload encryption.}
\label{fig:tls_encryption}
\end{figure}

\bheading{Record layer protocol.}
The record protocol of TLS protects the confidentiality and integrity of the
payload via symmetric encryption and MAC algorithms. Block encryption in the Cipher Block
Chaining Mode (CBC) is one of the most widely used modes of operation for block
ciphers. The encryption and authentication is conducted in the
\textit{MAC-pad-encrypt} scheme, as shown in \figref{fig:tls_encryption}. The
MAC of the data payload is first calculated to protect its integrity, and then
the resulting data is padded with dummy bytes (conforming to SSL/TLS
specifications) so that the total message size is multiples of the block size
(\eg, 16 bytes in AES). The resulting data blocks are then encrypted using the
symmetric cipher in the CBC mode. 

\subsection{Bleichenbacher Attacks against SSL/TLS}

Bleichenbacher attacks~\cite{bleichenbacher1998chosen} is the first practical
adaptive chosen-ciphertext attack against RSA cryptographic algorithms
conforming to the PKCS\#1 v1.5 encoding schemes. It exploits the format
correctness of the decrypted plaintext as an oracle and decrypts, by repeatedly
querying the oracle about the correctness of carefully-crafted ciphertexts, an
RSA public-key-encrypted message without the need of the RSA private keys.
Multiple studies have shown that Bleichenbacher Attacks have practical
implication in network security~\cite{bleichenbacher1998chosen,
manger2001chosen, klima2003attacking, bardou2012efficient}. Particularly, these
attacks have been demonstrated to work against SSL/TLS protocols that adopt RSA
algorithms to encrypt the \texttt{PreMasterSecrets} and at the same time reveal
non-conformant error messages over the network. Most widely used SSL/TLS
implementations today are believed to be immune to Bleichenbacher Attacks as
the oracle-enabling error messages have been suppressed. We have summarized
a brief history of the related studies in \secref{sec:related}.

%In this section, we describe the algorithm we implemented to conduct
%Bleichenbacher attacks against SSL/TLS protocols. Our attack is different from
%prior ones in that the format-conformant oracle is constructed using new
%side-channel analysis, but the algorithm to decrypt the plaintext, once the
%oracle is available, is similar to others.  Here we summarize the algorithm
%proposed by Bardou \etal~\cite{bardou2012efficient}, which is an optimized
%version of the original attacks~\cite{bleichenbacher1998chosen,
%klima2003attacking}. 

In this paper, we implemented the optimized Bleichenbacher attack proposed by
Bardou \etal~\cite{bardou2012efficient}. The attack relies on the artifact that
a correctly formatted \texttt{ClientKeyExchange} message, before encryption,
must begin with $0x0002$. Therefore, its value $m$ must satisfy $2B \leq m <
3B$, where $B$ is a constant that starts with $0x0001$ and ends with a running
of $8(k-2)$ of $0$s ($k$ is the length of $m$ in bytes).  By repeatedly querying
the oracle and finding a sequence of $s_i$ so that each $m_i$ ($c_i \equiv {c
\cdot (s_i)^e} \text{mod} \ n$, $m_i \equiv {(c_i)^d} \text{mod} \ n$ and $m
\equiv c^d \text{mod} \ n$) is also PKCS conformant, the adversary can gradually
narrow down the possible value ranges for $m$ until only one possibility
remains.  Interested readers can refer to Bardou
\etal~\cite{bardou2012efficient} for details of the algorithm. 

The oracle strength is defined as the conditional probability of the oracle
returning true given the decrypted plaintext message indeed begins with
$0x0002$.  A strong oracle means the adversary can complete the attack with
fewer queries. This is because the probability that a query with message $m$
beginning with $0x0002$ is returned by the oracle as false is smaller; hence the
adversary can collect a sequence of such messages faster.  For example, if the
oracle always returns true when the plaintext message starts with $0x0002$, the
oracle strength is 1.  In contrast, if the oracle returns true if and only if
the following three conditions hold at the same time, for instance, (1) the
first two bytes of the plaintext are $0x0002$, (2) the next 8 bytes do not
contain $0x00$, and (3) the next 246 bytes that follow (in the case of a
2048-bit RSA encryption) do have at least one byte of $0x00$, the oracle
strength is $${(\frac{255}{256})}^8 \times (1 - {(\frac{255}{256})}^{246})
\approx 0.599$$.

\subsection{CBC Padding Oracle Attacks}

It is known that the \textit{MAC-pad-encrypt} used in SSL/TLS is vulnerable to
padding oracle attacks~\cite{vaudenay2002security, Canvel:2003:PIS,
Fardan:2013:lucky, albrecht2016lucky, Moller:2014:poodle}. The vulnerability can
be exploited when the CBC mode of operation is chosen. In such attacks, a
carefully-crafted \textit{application data} packet is repeatedly sent by the
man-in-the-middle attacker to the vulnerable SSL/TLS server/client (collectively
called the SSL agent).  Each time the message is sent to the victim SSL agent
for decryption, the attacker modifies the ciphertext slightly to conduct an
\textit{adaptively chosen ciphertext attack}. The SSL agent checks the
correctness of the padding and the MAC after decrypting the message. If there
are errors in the format of the padding or content of the MAC, an alert message
will be returned. If the attacker can tell if the error message is caused by
only the padding error or by both padding errors and MAC errors, she has a
padding oracle that tells her whether her modification of the ciphertext is
decrypted into a correct padding (very unlikely to have a correct MAC). In SSL
v3.0, the padding format only specifies the last padding byte has the value of
the total padding length, while the other bytes could have random values. In TLS
specifications, all padding bytes have the same value, which is the number 
of the padding bytes. Therefore, a correct padding reveals the content of the last
byte of the plaintext, which gives the attacker the power to decrypt some data
without having the decryption key.

\begin{figure}[tbh]
\centering
\includegraphics[width=0.85\columnwidth]{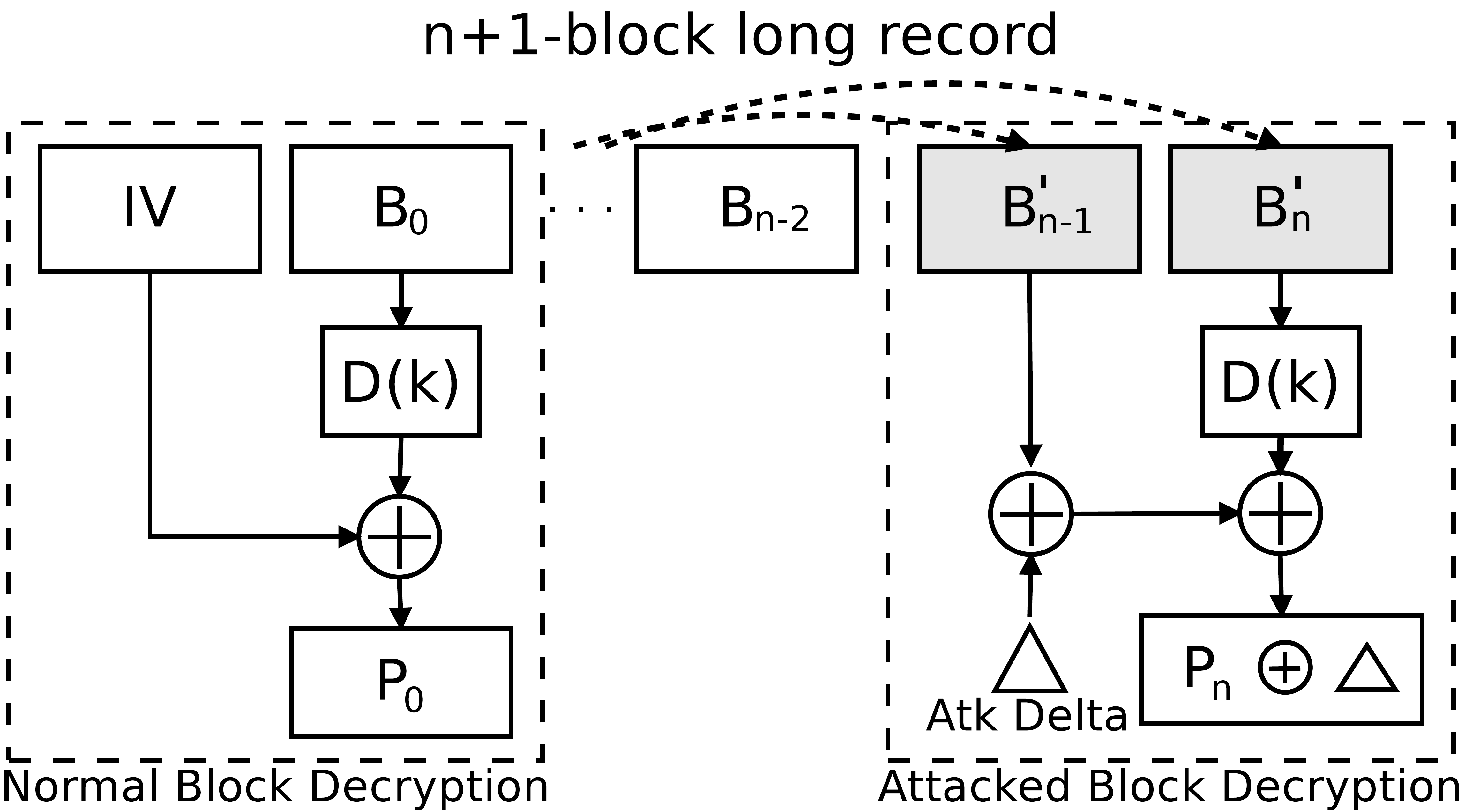}
\caption{Illustration of CBC padding oracle attacks.}
\label{fig:cbc_padding}
\end{figure}

More specifically, as shown in \figref{fig:cbc_padding}, the attacker can send a
ciphertext of $n+1$ blocks to the victim, with the last two blocks replaced by
the two blocks of interest, \eg, $B'_{n-1}$ and $B'_n$ in the figure.  Let's
denote the plaintext of $B'_n$ as $P'_n$, which is the value the attacker hopes
to learn. Therefore, $P'_n = B'_{n-1} \oplus D_k(B'_{n})$, where $D_k()$ is the
decryption function of a block under the secret key, $k$.  The attack proceeds
from the bytes on the right in the block and gradually extends to bytes on the
left: To decrypt the last two bytes of $P'_n$, the attacker, at each time she
queries the padding oracle, XORs the last two bytes of $B'_{n-1}$ with a value
$\Delta$, which has a value ranging from $0x0000$ to $0xFFFF$. Then the
resulting plaintext is $B'_{n-1} \oplus \Delta \oplus D(B'_n) = (B'_{n-1} \oplus
D(B'_n)) \oplus \Delta = P'_n \oplus \Delta$. When enumerating values of
$\Delta$ from $0x0000$ to $0xFFFF$, one of the values will lead to a correct padding
(\ie, $0x0101$ as the last two bytes). Therefore, the plaintext of the last two
bytes of $P'_n$ is simply $0x0101 \oplus \Delta$. The attack continues to guess
the other bytes on the right by altering the value of $\Delta$ and looking for a
correct padding of $0x020202$, $0x03030303$, \etc. 

The key to the success of such attacks is the ability to differentiate the cases
where a MAC error and a padding error occur at the same time from the cases
where only the padding error happens. When the error type is not reported, which
is the case in all SSL/TLS implementations after the publication of the original
padding oracle attacks in 2002~\cite{vaudenay2002security}, the CBC padding
oracle attacks becomes very difficult. The recently published variants of the
attack worked around this defense to re-enable the oracles: In Lucky Thirteen
attacks~\cite{Fardan:2013:lucky} and Lucky Microseconds
attacks~\cite{albrecht2016lucky}, a remote timing-channel orcale was used to
differentiate the two types of errors. In POODLE
attacks~\cite{Moller:2014:poodle}, the oracle is created by eliminating the MAC
errors when the plaintext is carefully crafted so that the length of the padding
is exactly one block (\ie, 16 bytes for AES). As such, only padding errors may
occur, which can be exploited as an oracle. This attack only works for SSL v3
because a correct padding only requires the last byte to be $16$, while
other padding bytes are not specified. In contrast, because the TLS protocols
specify the content of every padding byte to be the length of the padding, it
is unlikely to decrypt an arbitrary ciphertext into the correct padding. So the
attack won't work with TLS protocols.  In this paper, the oracle is
reconstructed using a new type of side channels and the attacks work on all
SSL/TLS versions and implementations.

% section threat model
\section{Threat Model Analysis}
\label{sec:threat}

To analyze the security threats on Intel SGX imposed by side-channel
attacks, in this paper, we systematically study one important category of
side-channel attacks---\attackNames. In these attacks, the goal is to
infer, by measuring side-channel observations, the indirect control-flow
transfers of the enclave program and thereby learning sensitive information that
is shielded by SGX. 

Inferring program control flows and learning sensitive information
are two separate steps. On one hand, the existence of side-channel attack
vectors, \eg, page-fault traces, cacheline access traces, branch instruction
traces, \etc., enables control-flow inference. We name such attacks \attackNames.
These attacks have been studied in previous studies. Here in this paper, we
propose a systematic approach to model \attackNames, which enables discussion of
these attacks \textit{without specifying the exact attack techniques}. On the other 
hand, control-flow leakage does not always lead to a security breach. Only code with
secret-dependent control flows is vulnerable to \attackNames. 
%We call such a vulnerability the \vulNames.

\subsection{\AttackNames}

\begin{figure}[t]
\centering
\includegraphics[width=0.85\columnwidth]{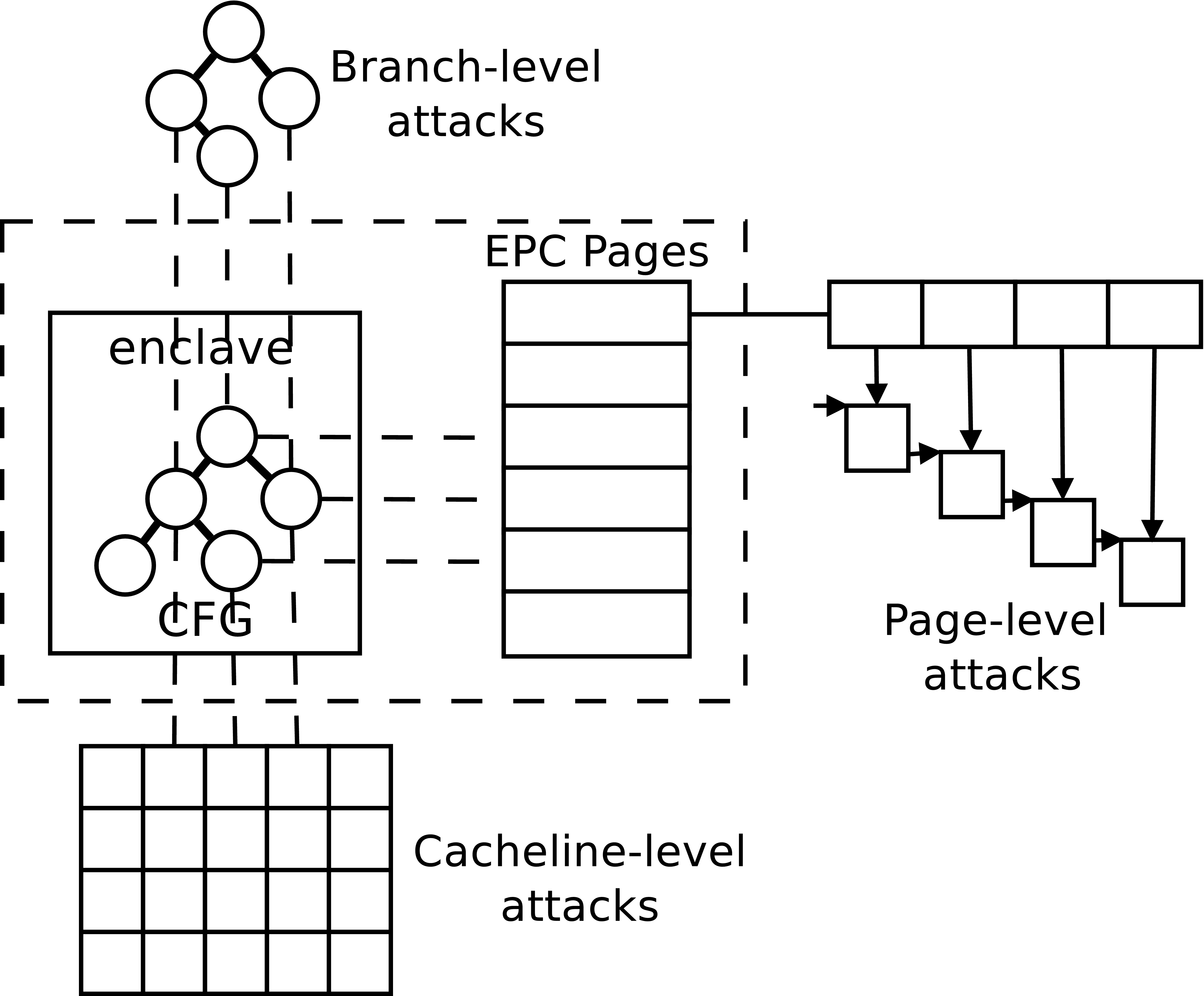}
\caption{Three categories of \attackNames.}
\label{fig:threat}
\end{figure}

In previous work, it has been shown that SGX enclaves are vulnerable to a
variety of side-channel attacks. Some manipulate page table entries,
some control shared caches, and others exploit shared branch prediction units.
The methods to collect side-channel observations are also diverse: Some attacks
use hardware timestamp counters to measure the execution time of specific code, some use Last
Branch Record (LBR) to measure elapsed cycles between branch instructions, and
some rely on deterministic events (\eg, page faults).  Regardless of the attack
techniques, we categorize \attackNames into three levels: page-level attacks,
cacheline-level attacks, and branch-level attacks. We illustrate these three
levels in \figref{fig:threat}.

\bheading{Page-level attacks.}
It was reported by Xu \etal~\cite{Xu:2015:controlled} that by clearing the
\textit{Present} bit of the page table entries, the adversary controlling the OS
can force the EPC page accesses by the enclave program to raise page fault
exceptions and be trapped into the OS kernel controlled by the adversary. In
this way, the adversary could observe the enclave program's page-level memory
access pattern\ignore{\cite{Xu:2015:controlled} states that at least one EPC
page needs to be allowed to access without page faults to avoid system halt.
Thus memory access patterns within the same page is not observable.}. In our own
exploration, we found not only the \textit{Present} bit, other bits in the page
table entry, such as \textit{Reserved} bits, \textit{NX} bit, \etc., as well as
Translation-Lookaside Buffers (TLB) and paging-structure
caches~\cite{IntelDevelopmentManual} also enable similar attack semantics. In
this paper, we model the side-channel observations collected in page-level
attacks as a sequence of page faults: $<P_1, P_2, P_3, \cdots, P_n>$, where
$P_i$ is the virtual page frame number of the enclave program. With known binary
code of the enclave programs, $P_i$ maps to a specific page of executable code
of the enclave program.  \ignore{If we allow $P_i = P_{i+1}$, that is,
consecutive accesses to the same page can be observed, we call the page fault
sequence \textit{consecutively duplicable sequence}. Otherwise, we call such a
sequence \textit{consecutively non-duplicable sequence}.}

\bheading{Cacheline-level attacks.} 
Intel SGX does not prevent cache-based side-channel attacks. Therefore, most
prior work on cache-based side-channel attacks is, in theory, applicable to SGX
enclaves. 
%These include \primeprobe attacks~\cite{liu2015practical, Inci:2015:seriously,
%Oren:2015:SSP}, \flushreload attacks~\cite{yarom2014flush, Zhang:2014:CSA,
%zhang2016return} on L1 data cache, L1 instruction cache, L2 cache, last-level
%caches (\aka, L3 caches). 
While it is challenging to model every single attack technique that has been
explored in previous studies, we abstractly model cache-based side-channel
attacks as a sequence of observations of the victim enclave program's cacheline
accesses: $<C_1, C_2, C_3, \cdots, C_n>$, where $C_i$ is the virtual address of
the beginning of the cacheline (\ie, cacheline sized and aligned memory block).
With known binary code of the enclave programs, $C_i$ maps to a specific
cacheline-sized block of executable code of the enclave program. \ignore{ Similar to
page-level attacks, if we allow $C_i = C_{i+1}$, that is, consecutive accesses
to the same cache-line can be observed, we call the cache-line access sequence
\textit{consecutively duplicable sequence}. Otherwise, we call such a sequence
\textit{consecutively non-duplicable sequence}.}

\bheading{Branch-level attacks.}
Very recently, Lee \etal~\cite{Lee:2016:SGXbranch} demonstrated that the control
flow of enclave programs can be precisely traced at every branch instruction
because the Branch Prediction Units (BPU) inside the CPU core is not flushed
upon Asynchronous Enclave Exit (AEX). Therefore, a powerful adversary could
interrupt the enclave execution, which triggers an AEX, and then execute a piece
of shadow code whose virtual addresses are the same as the victim code in the
lower 32-bit range---so that they are mapped to the same entries in the Branch
Target Buffer (BTB). The adversary employs LBR to learn whether each branch of
the shadow code is correctly predicted or not, which apparently is influenced by
the branch history of the enclave program that is just interrupted.  To model
these attacks, or other powerful attacks that are yet to be discovered, we
consider the strongest \attackNames as a sequence of basic blocks that are
executed in order: $<B_1, B_2, B_3, \cdots, B_n>$, where $B_i$ is a basic block
in the enclave program's control-flow graph (CFG). 

\subsection{\VulNames}

If the enclave program has secret-dependent control flows, then it is
potentially vulnerable to \attackNames. Such vulnerabilities are named \vulNames
in this paper. In this work, our focus is one of the most critical applications
for SGX enclave---SSL/TLS libraries.  Although SSL/TLS libraries, \eg, OpenSSL,
are implemented in a way that constant-time execution is enforced, however, as
we will show in this work, they still have \vulNames due to improper error
handling and reporting, thus are vulnerable to
\attackNames.

\ignore{
We assume the certificate validation process is secured by means of certificate
pinning or other similarly effective approaches. Otherwise, the malicious kernel
can easily spoof the DNS resolution results or the IP address of the
communicating parting to the enclave-protected SSL/TLS implementation, bypassing
the security check in the certificate validation.
}

% section vulnerability scanning tools
\section{Detecting SSL/TLS Vulnerabilities with \frameName}
\label{sec:detect}

In this section, we present the Side-channel Trace Analyzer for finding
Chosen-Ciphertext Oracles (\frameName), a differential analysis framework for
detecting \vulNames in SSL/TLS implementations under the threat model we laid
out in \secref{sec:threat}. The core idea behind the framework is that when
provided with encrypted SSL/TLS packets with non-conformant formats or incorrect
paddings with different types of errors, the decryption code may exhibit
different control flows that give rise to the decryption chosen-ciphertext
oracles. To enable automated tests for \textit{multiple} oracle vulnerabilities
on \textit{various} SSL/TLS implementations under \textit{different} attack
models, \ie, page-level, cacheline-level and branch-level \attackNames, we
developed a differential analysis framework (\secref{subsec:framework}) and used
it to evaluate 5 popular SSL/TLS libraries (\secref{subsec:results}).

\subsection{Differential Analysis Framework}
\label{subsec:framework}

\begin{figure}[t]
\centering
\includegraphics[width=0.98\columnwidth]{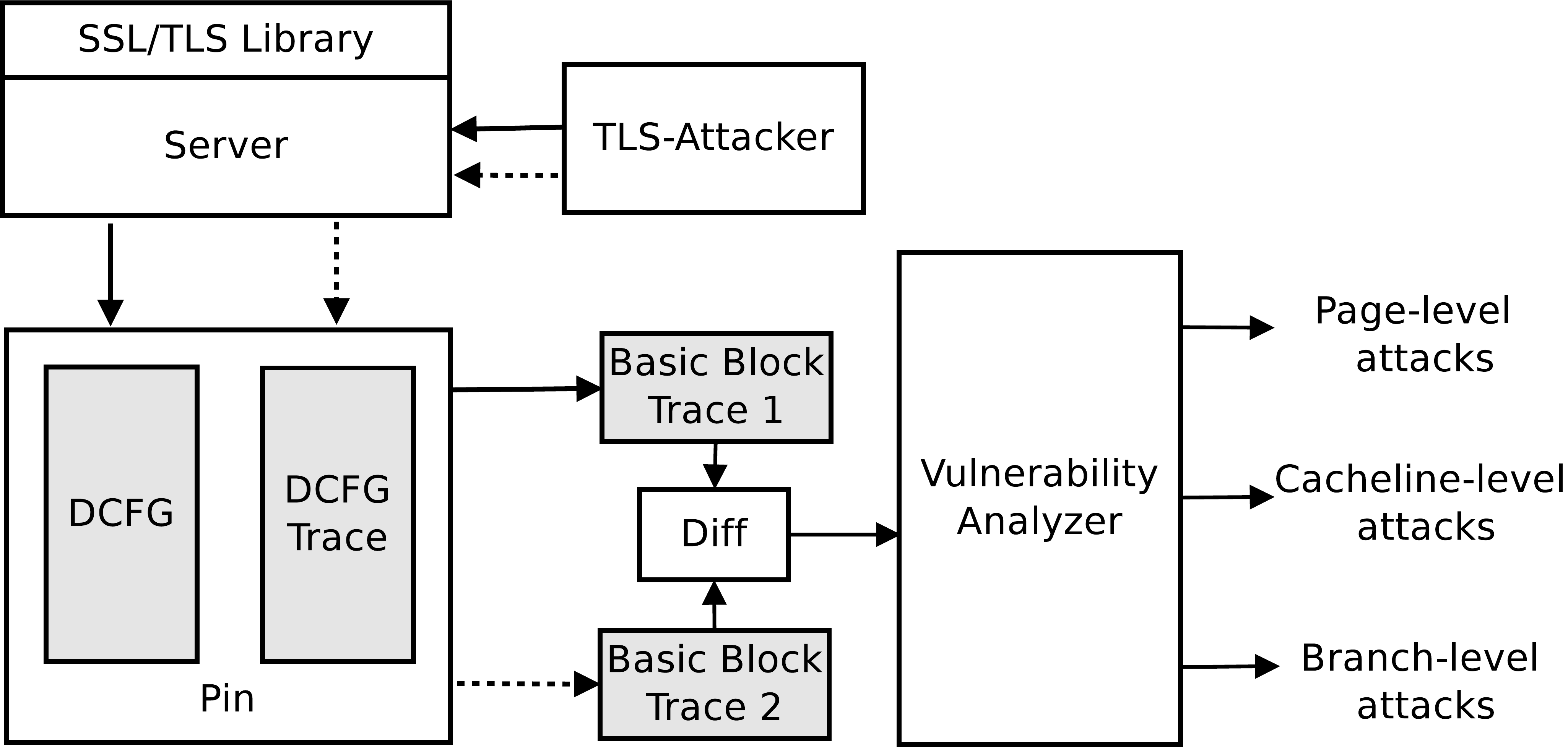}
\caption{Architecture of the differential analysis framework.}
\label{fig:architecture}
\end{figure}

At the center of our differential analysis framework is a dynamic
instrumentation engine to collect execution traces of the SSL/TLS
implementation.  The overall architecture of our framework is shown in
\figref{fig:architecture}.  Our framework consists of five components: a packet
generator (\ie, the TLS-attacker in the figure), an SSL/TLS program linked to an
SSL/TLS library under examination, a trace recorder (\ie, Pin), a trace
\textit{diff} tool, and a vulnerability analyzer.

A complete run of one differential analysis test follows \textit{three} main
steps. The \textit{first} step is to collect two execution traces.  The packet
generator generates two SSL/TLS packets following specific rules (to be
explained in \secref{subsec:results}) and sends them to the SSL/TLS program. The program
which is linked to the library being analyzed runs on top of the Pin-based trace
recorder, where the execution traces of the analyzed library are collected. The
\textit{second} step is to compare the two execution traces. Differences in the
traces indicate potential \vulNames. The \textit{final} step is to decide
whether the differences in the traces are exploitable by the attacker. Given a
specific attack model, \eg, page-level, cacheline-level, and branch-level
\attackNames, the vulnerability analyzer is able to tell whether the tested
library is vulnerable to such attacks and, if so, pinpoints the exploitable
vulnerabilities.

\bheading{Packet generator.}
The packet generator is in charge of generating the input to the framework. It
prepares encrypted packets with specified plaintext or ciphertext (with
specified errors) to be sent to the peer at any stage of an SSL/TLS connection.
In our implementation, we adopted an open-source tool, 
TLS-attacker~\cite{somorovsky2016systematic}. It
is able to complete an SSL/TLS handshake or replace any packet in this
process.  It is also able to send arbitrary data records after the SSL/TLS
connection has been successfully established. 

\bheading{Trace recorder.}
A core component of the framework is the execution trace recorder. We
implemented the trace recorder on top of Intel Pin. Pin~\cite{luk2005pin} 
is a dynamic binary instrumentation framework
that is suitable for a range of program analysis tasks. It enables various
tools, called \textit{Pintools}, to be developed using the framework. Of
interest to our purpose is its capability of dynamic instrumenting a software
program \textit{without changing its memory layout}, which is essential for
detecting \vulNames. Particularly, a Pintool provided by 
Pinplay kit~\cite{patil2010pinplay} can be used to create the Dynamic 
Control-Flow graph (DCFG) of a program~\cite{yount2015graph}. 

\begin{figure}[tbh]
\centering
\includegraphics[width=0.98\columnwidth]{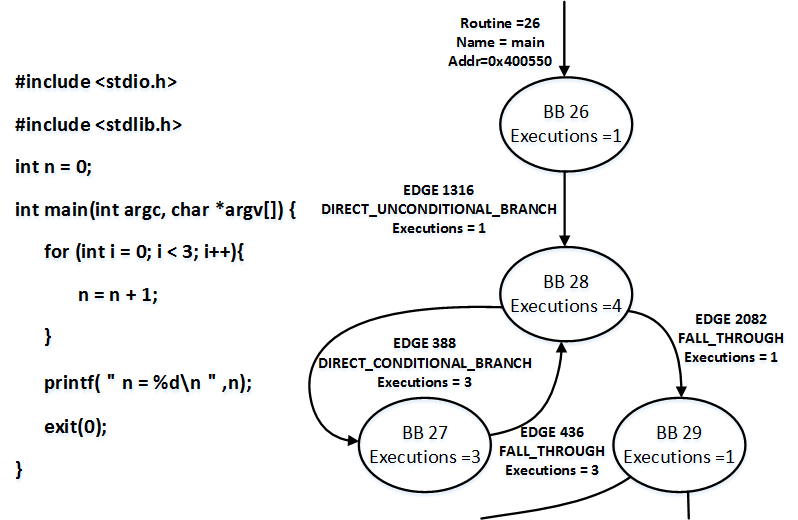}
\caption{An example of DCFG.}
\label{fig:dcfg}
\end{figure}

DCFG is defined by Intel as an extension of the control-flow graphs
(CFG)~\cite{allen1970control}.  An example of a DCFG can be found in
\figref{fig:dcfg}. Generally speaking, a DCFG shows the portion of a CFG that
has been executed.  An edge in a DCFG is augmented with a counter, which records
the number of times this edge is executed. Pinplay kit also provides an option
to record the exact sequence of the executed edges in the DCFG, which is called
DCFG-Trace.  Combining the DCFG with the DCFG-Trace, \frameName is able to
generate a trace of basic blocks that has been executed by the instrumented
programs.

In order to improve the runtime performance and to facilitate data analysis, we
need to specify which parts of the execution we are most interested in. For
example, if we are looking for vulnerabilities in the handshake protocol, the
execution trace should be recorded only when the handshake APIs are called.
However, we found that such a selective tracing functionality that could have
been enabled by control options \texttt{enter\_func} and \texttt{exit\_func} in
the Pintool could not work properly with the SSL/TLS libraries. As a solution,
we added in the SSL/TLS program two empty functions \texttt{foo()} and
\texttt{bar()} to wrap the functions that we are interested in, by adding a call
to \texttt{foo()} before it and \texttt{bar()} after it. Thus we could control
the Pintool to selectively trace functions when the option \texttt{-log:control
start:address:foo,stop:address:bar} is enabled.

The output of the trace recorder consists of two \texttt{JSON} files. One
includes the DCFG as well as basic information such as base addresses of the
libraries and the offsets of each of the basic blocks in the libraries. The
other \texttt{JSON} file contains a trace of DCFG edges. We then extended the
Pintool using Pin DCFG APIs to merge the two files into a trace of basic blocks.
Because the base addresses of the libraries change every time we run the
program, we use the name of the library and the offset of the basic block to
uniquely identify a basic block.

It might be worthwhile noting here, that Pin has bugs when executing certain
functions (\eg, \texttt{gnutls\_record\_recv()} in GnuTLS) and, as a result, the
DCFG traces of these functions cannot be correctly recorded. For this special case, we
replace the Pin-based trace recorder with a Callgrind-based trace recorder.
Callgrind is a tool of Valgrind~\cite{nethercote2007valgrind, weidendorfer2004tool} 
which is also a dynamic instrumentation
framework.  We extend Callgrind to include timestamps for each function call in
order to recover the complete call trace from the call graph generated by
Callgrind. Unfortunately, Callgrind does not provide fine-grained basic block
tracing (\eg, DCFGs) like Pin.

\bheading{The diff tools.}
Because the Pin-generated basic-block traces are ordered sequences of basic
blocks, the \textit{diff} command of Linux OS turns out to be enough to identify
the differences between the two traces.  In contrast, the Callgrind-generated
function call traces are less structured.  We, therefore, implemented a Python
tool based on \textit{difflib} to compare Callgrind-generated function call
traces, which first converts the call traces to call trees with nodes associated
with timestamps, and then represent the call trees using XML, which can be
compared using \textit{difflib}.

\bheading{Vulnerability analyzer.}
Given the results of the \textit{diff} tools, we build a vulnerability analyzer
in Python to examine \vulNames. Particularly, the differences in the basic-block
traces are by themselves vulnerability to branch-level attacks. To detect
vulnerabilities at the cacheline level or the page level, we need to convert the
basic-block traces into traces of cachelines and pages. Specifically, the
virtual address of the beginning of each basic block is calculated. 
%Although the
%real virtual address changes at each run due to address space layout
%randomization, the portion of the virtual address that matters to our analysis is
%the offset within a binary, which does not change given the binary code. We
%artificially assign a unique, large virtual address to the beginning of the
%libraries and the binaries, so that each basic block has a unique virtual address.
The corresponding page trace can be obtained by
dividing the virtual addresses of the basic blocks by 4096, the size of a memory
page, and then merge consecutive basic blocks together if they have the same page address.  
Similarly, cacheline traces can be generated by dividing virtual addresses of
the basic blocks by 64, which is the size of a cacheline.
After the conversion, if the trace differences in the basic-block sequences render 
the same cacheline trace or page trace, the program, as per this test, is not 
vulnerable to cacheline-level or page-level \attackNames.

\subsection{Evaluation and Results}
\label{subsec:results}

\begin{table*}[t]
\begin{minipage}{1.7\columnwidth}
\centering
\caption{Experiment results of the differential analyses. B: vulnerable to
branch-level attacks? C: vulnerable to cacheline-level attacks? P: vulnerable to
page-level attacks?  D: Differentiable; N: Not differentiable; N/A:
unable to test.}
\label{tab:detection}
\end{minipage} \\[-10pt]

\centering \small{
\begin{tabular}[t]{c|c|ccc|ccc|ccc|ccc|ccc}
\Xhline{1pt}
&
\multirow{3}{*}{\textbf{Test Name}} & 
\multicolumn{3}{c|}{\textbf{OpenSSL}} &
\multicolumn{3}{c|}{\textbf{GnuTLS}} & 
\multicolumn{3}{c|}{\textbf{mbedTLS}} & 
\multicolumn{3}{c|}{\textbf{WolfSSL}} & 
\multicolumn{3}{c}{\textbf{LibreSSL}} \\
& & 
\multicolumn{3}{c|}{\textbf{1.0.2j}} &
\multicolumn{3}{c|}{\textbf{3.4.17}} & 
\multicolumn{3}{c|}{\textbf{2.4.1}} & 
\multicolumn{3}{c|}{\textbf{3.10.0}} & 
\multicolumn{3}{c}{\textbf{2.5.0}} \\
&
& \textbf{B} & \textbf{C} & \textbf{P} 
& \textbf{B} & \textbf{C} & \textbf{P} 
& \textbf{B} & \textbf{C} & \textbf{P} 
& \textbf{B} & \textbf{C} & \textbf{P} 
& \textbf{B} & \textbf{C} & \textbf{P} 
\\
\Xhline{1pt}
\multirow{10}{*}{\rotatebox[origin=c]{90}{\parbox{8em}{\centering Bleichenbacher
attacks}}} &
{PKCS\#1 Conformant}  & D & D & D & D & D & D & D & D & D & D & D & D & D & D & D \\
& {Wrong Version}  & D & D & D & D & D & D & D & D & D & D & D & N & D & D & D \\
& {No 0x00 Byte}  & D & D & N & D & D & D & D & D & D & D & D & N & D & D & N \\
& {0x00 in Padding}  & D & D & D & D & D & D & D & D & D & D & D & N & D & D & D \\
& {0x00 in PKCS Padding}  & D & D & N & D & D & D & D & D & D & D & D & D & D & D & N \\
& {PMS Size=0}  & D & D & D & D & D & D & D & D & D & D & D & N & D & D & D \\
& {PMS Size=2}  & D & D & D & D & D & D & D & D & D & D & D & N & D & D & D \\
& {PMS Size=8}  & D & D & D & D & D & D & D & D & D & D & D & N & D & D & D \\
& {PMS Size=16}  & D & D & D & D & D & D & D & D & D & D & D & N & D & D & D \\
& {PMS Size=32}  & D & D & D & D & D & D & D & D & D & D & D & N & D & D & D \\
& Exploitable & \cmark & \cmark & \cmark & \cmark & \cmark &
\cmark & \cmark & \cmark & \cmark & \cmark & \cmark &
\cmark & \cmark & \cmark & \cmark \\
\Xhline{1pt}
\multirow{7}{*}{\rotatebox[origin=c]{90}{\parbox{7em}{\centering Padding Oracle
attacks}}} &
{Padding Length Byte XOR 1}  & D & D & N & N/A & N/A & D & D & D & D & D & D & D & D & D & D \\
& {Padding Length Byte = 0x00}  & D & D & N & N/A & N/A & D & D & D & D & D & D & D & D & D & D \\
& {Padding Length Byte = 0xFF}  & D & D & N & N/A & N/A & D & D & D & D & D & D & D & D & D & D \\
& {Last Padding Byte XOR 1}  & D & D & N & N/A & N/A & D & D & D & D & D & D & D & D & D & D \\
& {Last Padding Byte = 0x00}  & D & D & N & N/A & N/A & D & D & D & D & D & D & D & D & D & D \\
& {Last Padding Byte = 0xFF}  & D & D & N & N/A & N/A & D & D & D & D & D & D & D & D & D & D \\
& Exploitable & \cmark & \cmark & \xmark & N/A & N/A & \cmark &
\cmark & \cmark & \cmark & \cmark & \cmark & \cmark &
\cmark & \cmark & \cmark \\
\Xhline{1pt}
\end{tabular}} \\%[4pt]

\end{table*}

We applied \frameName to detect two types of oracle
attacks, CBC padding oracle attacks, and Bleichenbacher attacks, in the latest
versions (as of February 2017) of five popular open-source libraries (see
\tabref{tab:detection})\footnote{As of August 2017, Intel's SGX SDK~\cite{intelsgxssl}
only contained a cryptographic library; and the SSL/TLS implementation was not
completed. Therefore, we could not conduct tests on Intel's official SGX SSL/TLS
implementation.}.    

\subsubsection{Bleichenbacher Tests}

To detect vulnerabilities that enable Bleichenbacher attacks, we conducted a
series of differential tests. In each of the tests, we differentially analyzed
two variations of the \texttt{ClientKeyExchange} messages in the handshake
protocol. One of the two variations is a non-conformant message in which the
first two bytes of the plaintext is not $0x0002$, which we call the
``\texttt{standard error}''; the other variation is a message following one of
the ten rules specified in \tabref{tab:detection}. For example,
``\texttt{PKCS\#1
conformant}'' means that the \texttt{ClientKeyExchange} message is correctly
formatted according to PKCS\#1 standard, but with an incorrect PMS. If this
message is not differentiable from the ``\texttt{standard error}'', the library
is not vulnerable to Bleichenbacher attacks. ``\texttt{Wrong Version}'' stands
for the two version-number bytes are incorrect. In the ``\texttt{No 0x00 Byte}''
test case, the delimiter $0x00$ byte after the padding bytes is changed into a
non-$0x00$ byte. The test cases ``\texttt{0x00 in PKCS Padding}'' and
``\texttt{0x00 in Padding}'' mean that some bytes in the corresponding padding
bytes are modified to $0x00$, which they should not if the messages are
conformant. Note that the first 8 bytes of the padding string are the PKCS
paddings while the rest bytes are regular paddings.  They are treated by the
SSL/TLS library differently.  ``\texttt{PMS Size}'' test cases are performed by
moving the $0x00$ byte to somewhere in the middle of the
\texttt{PreMasterSecret} (PMS) so that PMS is truncated. For example,
``\texttt{PMS Size=2}'' is done by moving $0x00$ to the third last byte.  Note
all the \texttt{PreMasterSecret} (PMS) in these tests are invalid so that the
error handling procedure is always triggered.

The results of the analysis are shown in \tabref{tab:detection}. ``\texttt{D}''
suggests that in the differential analysis, the two traces, when converted to
the corresponding level, are differentiable; ``\texttt{N}'' means the two traces
are not differentiable.  If ``\texttt{PKCS\#1 conformant}'' is differentiable
from ``\texttt{standard error}'', it means we can construct an oracle that when
it returns true, we are certain that the corresponding plaintext message starts
with $0x0002$. This means the tested library is considered exploitable by a
Bleichenbacher attack (labeled ``\texttt{\cmark}'' in the row with the header
``\texttt{Exploitable}''). However, what we do not know is whether the oracle
returning false means the message does not begin with $0x0002$. If, at the same
time, some of the other 9 tests yield differentiable traces, it means we have a
higher probability to assert that when the oracle returns false the message does
not start with $0x0002$, which leads to a stronger oracle.

In the cacheline-level and branch-level \attackNames, the oracle
strength is 1 for all libraries; this is because the oracle only returns false
when ``\texttt{standard error}'' happens. In these cases, we have a very strong
oracle that can help break the secret with fewer queries. The page-level attacks
against GnuTLS and mbedTLS also have an oracle strength of 1, but those for
OpenSSL and LibreSSL are lower, which is roughly $(\frac{255}{256})^8 \times (1
- {(\frac{255}{256})}^{49})  \approx 0.1691$ when the RSA key
size is 2048 (see \figref{fig:blockformat})\footnote{It requires $0x00$ in the
  8-byte PKCS padding and No $0x00$ in the last $48+1$ bytes.}. It means the
adversary needs to send more (roughly, $\frac{1}{0.1691} = 5.9\times$) queries
to the ``weaker'' oracle compared to using a stronger oracle (\ie, oracle
strength is 1).  The oracle strength for page-level \attackNames against WolfSSL
is the lowest, because most of the differential tests render
\textit{non-differentiable}, making the oracle attack very slow.

\subsubsection{Padding Oracle Tests}

To detect vulnerabilities that give rise to CBC padding oracle attacks, we also
performed a series of tests. In each test, the framework is provided with two
\textit{application data} messages that are encrypted with symmetric keys in
the CBC mode. All messages are four blocks in length. One message only has an
incorrect MAC, the ``\texttt{standard error}'', and the other message has both
an incorrect MAC and one of the six padding errors listed in
\tabref{tab:detection}.  Specifically, the six test cases can be divided into
two groups: The first group of tests is conducted by modifying the
``\texttt{Padding Length Byte}'' which is the last byte of a record; the second
group modify the last ``\texttt{Padding Byte}'' which is the second last byte of
a record. The two groups each generate one error padding case by (1) XORing the
target byte with 1, (2) setting it to 0x00 and (3) setting it to 0xFF. We note
that the test of ``\texttt{Padding Length Byte = 0x00}'' is special. When it
yields differentiable traces, it means the padding length cannot be 0x00,
therefore our test should break the last two bytes together (by looking for
paddings of $0x0101$). Otherwise, the attack can start with guessing the last
byte, greatly reducing the complexity of the attacks. None of the SSL/TLS
implementations we tested allow padding length to be $0x00$.  But it only makes
the attack slightly longer and does not eliminate its vulnerability.

When the ``\texttt{standard error}'' can be differentiated from the various
padding errors, the library is vulnerable to padding oracle attacks.  
The results of the analysis are shown in \tabref{tab:detection}.  We can see
that almost all libraries, except for OpenSSL, are vulnerable to all levels of
\attackNames. OpenSSL 1.0.2j is not vulnerable to page-level attacks because
all its distinguishable traces are contained in the same page. Another
exception is GnuTLS. As we have mentioned before, the Pintool does not support
GnuTLS's \texttt{gnutls\_record\_recv()} function due to a bug in the tool.
Consequently, we used Callgrind-based trace recorder, which does not support analysis at
the branch level and the cacheline level. 

\ignore{
\begin{table}[t]
\centering \small{
\begin{tabular}[t]{c|c|c|c|c|c}
   & Wrong Version & No 0x00 Byte & 0x00 in Padding \\
\Xhline{1pt}
\textbf{$P(U_i)$} & $1 - {\frac{1}{256}}^2$ & $\frac{255}{256}$ & $1 - {\frac{1}{256}}^{197}$  \\
\Xhline{1pt}
   & 0x00 in PKCS Padding & PMS = 0 & $1\leq$PMS $\leq2$ \\
\Xhline{1pt}
\textbf{$P(U_i)$} & $1 - {\frac{1}{256}}^8$ & $\frac{1}{256}$ & $1 - {\frac{255}{256}}^2$  \\
\Xhline{1pt}
   & $3\leq$PMS $\leq8$ & $9\leq$PMS $\leq16$ & $17\leq$PMS $\leq32$\\
\Xhline{1pt}
\textbf{$P(U_i)$} & $1 - {\frac{255}{256}}^6$ & $1 - {\frac{255}{256}}^8$ & $1 - {\frac{255}{256}}^{16}$  \\
\end{tabular}} \\[4pt]
\begin{minipage}{0.8\columnwidth}
\centering
\caption{Probability of Undistinguishable tests.}
\label{tab:probability}
\end{minipage} \\[-10pt]
\end{table} 
}

\subsubsection{Findings}
Although some cryptographic libraries, such as OpenSSL, aim to enforce
constant-time implementations, \frameName finds that all five SSL/TLS libraries
are vulnerable to \attackNames. In most cases, page-level attacks are sufficient
to create an oracle and perform oracle attacks against these libraries. We
scrutinized the identified vulnerabilities and found that there are primarily
two reasons for the leakage. First, the oracles for Bleichenbacher attacks are
typically caused by the improper error logging and reporting mechanisms in the
library. Second, the oracles for padding oracle attacks are typically created by
the improper implementation of constant-time cryptography in the patches for the
existing side-channel attacks (\eg, the Lucky Thirteen attack, cache attacks,
\etc.).  We briefly summarize one example in this section, and list vulnerable code for
other vulnerabilities in \appref{app:examples}. 

%In the next section, we will demonstrate several oracle attacks against these vulnerable SSL/TLS implementations.  

Particularly, the CBC padding oracle in GnuTLS v3.4.17 can be constructed by
monitoring the execution order of the function
\texttt{ciphertext\_to\_compressed()} (see Listing~\ref{lst:gnutls:cbc1}) and
the function \texttt{\_gnutls\_auth\_cipher\_add\_auth()}
(Listing~\ref{lst:gnutls:cbc2}).  Specifically, \texttt{dummy\_wait()} is called
in \texttt{ciphertext\_to\_compressed()} when the padding or MAC is incorrect.
This function was designed to defeat the timing-based Lucky Thirteen
attack~\cite{Fardan:2013:lucky} by introducing intentional delays. However,
\texttt{dummy\_wait()} checks if the error is caused by incorrect padding (line
3 of Listing~\ref{lst:gnutls:cbc2}), and calls
\texttt{\_gnutls\_auth\_cipher\_add\_auth()} (line 8 and 12 of
Listing~\ref{lst:gnutls:cbc2}) if the padding is correct (and the MAC is
incorrect).  In this example, the decryption oracle is introduced by the defense
against timing attacks, but the control flow of the additional delay is leaked
to the more powerful man-in-the-kernel attackers.

\begin{figure}[tbh]
\centering
\begin{minipage}{.96\columnwidth}
\begin{lstlisting}[caption={Snippet of
\texttt{ciphertext\_to\_compressed()}.},
label={lst:gnutls:cbc1}]
	...

	ret =
	    _gnutls_auth_cipher_tag(&params->read.cipher_state, tag, tag_size);
	if (unlikely(ret < 0))
		return gnutls_assert_val(ret);

	if (unlikely
	    (gnutls_memcmp(tag, tag_ptr, tag_size) != 0 || pad_failed != 0)) {
		/* HMAC was not the same. */
		dummy_wait(params, compressed, pad_failed, pad,
			   length + preamble_size);

		return gnutls_assert_val(GNUTLS_E_DECRYPTION_FAILED);
	}

	...
\end{lstlisting}
\end{minipage} 
\vspace{-0pt}
\end{figure}

\begin{figure}[tbh]
\centering
\begin{minipage}{.96\columnwidth}
\begin{lstlisting}[caption={Snippet of
\texttt{dummy\_wait()}.},
label={lst:gnutls:cbc2}]
	...

	if (pad_failed == 0 && pad > 0) {
		len = _gnutls_mac_block_size(params->mac);
		if (len > 0) {
			if ((pad + total) % len > len - 9 && total % len <= len - 9) {
				if (len < plaintext->size)
					_gnutls_auth_cipher_add_auth
						(&params->read.cipher_state,
					     plaintext->data, len);
				else
					_gnutls_auth_cipher_add_auth
						(&params->read.cipher_state,
					     plaintext->data, plaintext->size);
			}
		}
	}
\end{lstlisting}
\end{minipage} 
\vspace{-0pt}
\end{figure}

% section result and case studies
\section{Vulnerability Validation}
\label{sec:attack}

To validate the detected \vulNames by \frameName, in
this section, we describe in details two types of oracle attacks against SSL/TLS
implementations in enclaves: CBC padding oracle attacks and Bleichenbacher
attacks. 

%These attacks are examples of the \textit{man-in-the-kernel} attacks. 

\subsection{Attack Implementation}
\label{subsec:implementation}

\begin{wrapfigure}{r}{0.5\columnwidth}
\centering
\includegraphics[width=\linewidth]{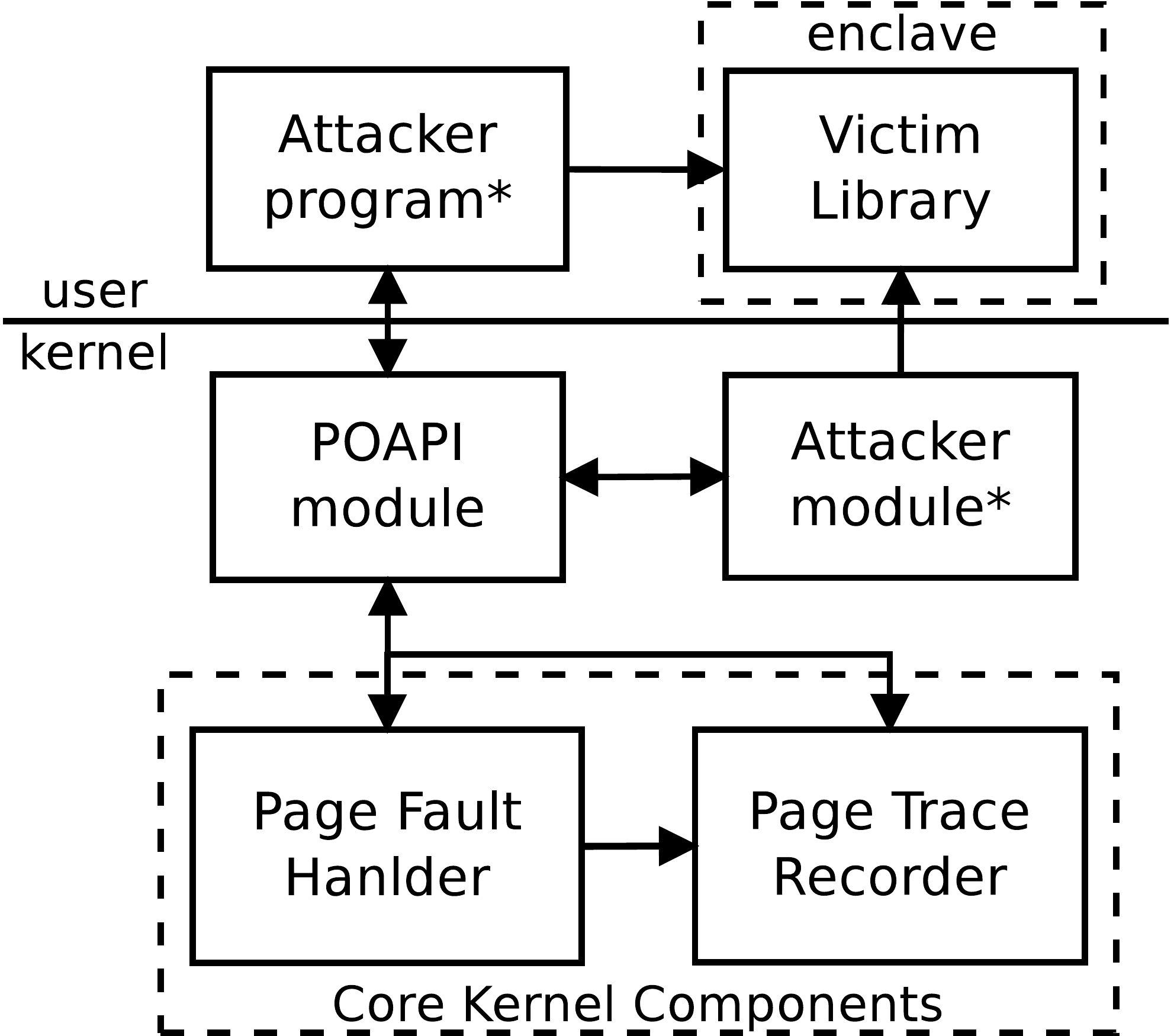}
\caption{Overview of the attack implementation.}
\label{fig:attack_overall}
%\end{figure}
\end{wrapfigure}

We implemented the page-level control-flow inference attacks by extending the
mainstream Linux operating system kernel. Our implementation of the attack is shown
in \figref{fig:attack_overall}. The extensions to the kernel include several
loadable kernel modules (LKM) and some modification of the core
kernel components.

In the core kernel components, particularly, we modified the page-fault handling
routine so that it is able to tackle a category of page faults
that are triggered because one of the reserved bits (\eg, bit 51 in our
implementation) in the page table entries (PTE) is set.  This reserved bit in
PTE is not already used by the Linux kernel, so only the attack code could
have triggered this type of faults.  When such page faults are intercepted, the
page-fault handler resets the reserved bit of the corresponding PTE to 0 so that
future accesses to the same page will be allowed (because otherwise the process
will hang due to frequent page faults); it also sets the reserved bit of
the last accessed page, tracked in a global variable in the kernel, to 1 in
order to capture future access of it. In sum, the kernel only allows one
executable page in the ELRANGE of the victim process (that the attacker is
interested in monitoring) to be accessible at a time. 

%\begin{figure}[tbh]

A data array, dubbed \textit{Page Trace Recorder (PTR)}, is added to the
kernel space for the page-fault handler to record the list of virtual pages that
has been accessed by the enclave program. Each time a page fault triggered by
the reserved bit in PTEs occurs, the faulting page is appended to the list,
which also increments a global counter by one. 

The attacks can be initiated either from the userspace or from the kernel. To
facilitate the attacks, we implemented a set of kernel interfaces, dubbed Paging
Oracle Attack Program Interface (POAPI), that can be triggered from both
userspace and kernel space.  The interfaces are encapsulated as a kernel module,
\ie, the POAPI module in \figref{fig:attack_overall}. The interfaces are either
used by a userspace program, the attacker program in the figure, or by another
kernel module, the attack module in the figure. As either an attack program or
an attack module is needed in the two attacks we describe shortly, they are
labeled with asterisks to indicate only one of them is needed in an attack. 

To initiate the attack, the POAPI module is provided with the name of the victim
process, the virtual addresses of the EPC pages to be monitored, and the
specific page sequence (specified using page indices rather than virtual addresses)
to be monitored for oracle construction.  The sequence of pages is also called
the \textit{template sequence}. POAPI locates the page tables of the victim
process in the kernel and sets the reserved bits of the PTEs to be 1 so that
accesses to these pages by the enclave code will be trapped into the kernel. The
template sequence is translated into the sequence of virtual pages in this step,
so it can be matched later with virtual page sequences in PTR.  POAPI provides
two addition interfaces: First, a \texttt{Reset()} call that will reset the PTR to
empty.  This functionality is important in our oracle attacks as we need to
repeatedly query the oracles. Second, an \texttt{Oracle()} call that will return
true or false: If the template sequence matches the entire sequence in PTR, then
\texttt{Oracle()} returns true; otherwise it returns false.

\subsection{CBC Padding Oracle Attacks}
\label{subsec:cbc}

\bheading{The oracle.}
We demonstrated the CBC padding oracle attack on the implementation of TLS v1.2
in GnuTLS 3.4.17 (latest version as of February 2017).  \frameName suggests
that this implementation is vulnerable to page-level control-flow inference
attacks. More specifically, the correctness of the paddings can be revealed by
the execution order of two functions, \texttt{ciphertext\_to\_compressed()} and
\texttt{\_gnutls\_auth\_cipher\_add\_auth()}.  We found in our experiments that
by monitoring only the memory pages that contain these two functions the
adversary is able to construct a powerful oracle for plaintext recovery.
Therefore, our template sequence only contains two memory pages. Note that
\texttt{ciphertext\_to\_compressed()} is large and spans two pages. We selected
the second page to monitor. This is instructed by the differential analysis tool
already, so no manual analysis is needed.  By labelling the memory page
containing \texttt{ciphertext\_to\_compressed()} as index 0 and that containing
\texttt{\_gnutls\_auth\_cipher\_add\_auth()} as index 1, the template sequence
is ``1010101010''.

\bheading{Detailed implementation.}
We run the victim library inside SGX with help of
Graphene-SGX~\cite{graphene-sgx}.  Particularly, the victim GnuTLS library we
attack is loaded as \texttt{sgx.trusted\_files} into enclaves with the victim
server programs.  However, GnuTLS does not support Intel SGX: The initialization
of the library will check the availability of accelerated encryption
instructions with the \texttt{CPUID} instruction---an instruction not supported
by SGX. 
%Since SGX does not support the accelerated encryption instructions anyway,
%there is no need to do the check at all. 
Thus we modified the library slightly by simply removing the check to allow it
to run directly in enclave (inside Graphene).

%\begin{figure}[t]
%\centering
%\includegraphics[width=0.85\columnwidth]{fig/cbc_workflow}
%\caption{Workflow of the CBC padding oracle attacks.}
%\label{fig:cbc_workflow}
%\end{figure}

The padding oracle attack is implemented as a kernel module that leverages the
POAPI to query the padding oracle constructed from the page-level control-flow
inference attacks. The attack starts after the SSL/TLS server in Graphene has
been launched. The process name, the virtual address of the two memory pages
that contain the two functions, and a template sequence ``1010101010'' are
provided through POAPI. If the encrypted message has a valid padding but invalid
MAC, \texttt{Oracle()} will find a match in the PTR and return true. If both the
padding and the MAC are invalid, a sequence of ``10101010'' will be found in the
PTR.  

Following prior studies~\cite{Fardan:2013:lucky, irazoqui2015lucky}, the padding
oracle attack is performed over multiple TLS sessions. This attack is practical
when the victim client can be triggered to repeatedly establish new TLS
connections with the victim server and send the same message in each new 
connection.  Particularly, the victim client first establishes a TLS connection
with the victim server using the SSL handshake protocol and negotiates to use the
\texttt{TLS\_RSA\_WITH\_AES\_128\_CBC\_SHA} ciphersuite in TLS v1.2 (through a
process that can be heavily influenced by the man-in-the-kernel attacker).  Then
it sends an encrypted data record to the server. The man-in-the-kernel attacker
modifies the ciphertext to prepare its query to the oracle.  If the server
receives a data record with incorrect MAC or incorrect padding, it sends a
\textit{bad\_record\_mac} alert to the client and shuts down the current TLS
session. When the client receives the alert, it immediately restarts a new TLS
connection to start a new query. The ciphertext will be different from the first
time, as the symmetric key, \ie, AES key, used to encrypt the data is different.
The adversary will intercept the message, again, and make modifications
according to its adaptive query strategies.

The attack kernel module we implemented for the CBC padding oracle attacks, upon
kernel module initialization, also registers a Netfilter to intercept all the
traffic sent to and from the server process, by filtering traffic with
specific port number. More specifically, two hooks were registered with hooknum of
\texttt{NF\_IP\_LOCAL\_IN} and \texttt{NF\_IP\_LOCAL\_OUT}.  With this
functionality, the adversary is able to examine each SSL/TLS packet and
determine the packet type by reading the first five bytes in the data segment of
packets. Byte 1 indicates the content type.  The adversary is interested in two
particular types: $0x15$ and $0x17$. $0x15$ means the packet is an Alert message and
$0x17$ means Application Data. Byte 2 and 3 are TLS versions. Since we are
attacking TLS 1.2, they should be $0x0303$. The last two bytes indicate the
compressed plaintext length. If an Alert message sent from the SSL/TLS server to
the client is observed by the kernel module, it means that the server has
decrypted the mal-formatted record and sent the client a
\texttt{bad\_record\_mac} alert, the adversary immediately checks the whether
the corresponding plaintext padding of the modified record is valid by calling
\texttt{Oracle()}. Notice that when the Netfilter intercepts the packet and modifies 
the ciphertext, all the \texttt{checksum}s, such as IP and TCP \texttt{checksum}s, 
will fail when checked by the kernel. Thus, a special flag is added to the 
modified packets and the kernel is modified to bypass all packet integrity 
checks upon appearance of this flag.

\ignore{If the padding is
correct, the plaintext of the cracked byte can be inferred by XORing the
expected padding with the first byte of $\Delta$. For the last two bytes, their
plaintext should be 0x01 01 XOR $\Delta$ ($\Delta$ is two-byte long). And when
intercepting the next incoming Application Data packet, the attacker could start
recovering the byte ahead of the cracked byte. If the oracle shows that the
padding is incorrect, the attacker will set the first byte (or the two bytes
when cracking the last two bytes) of $\Delta$ to another value and try with the
next incoming Application Data packet.}

\bheading{Evaluation.}
The complexity of plaintext recovery with AES encryption is at most  $2^{16} +
14 \times 2^8 = 69,120$ queries. This is because the last two bytes need to be
enumerated together, but the rest of the
bytes can simply be decrypted one byte after another, leading to a linear
complexity in the size of the block.  In our experiment to decrypt one block
with random data, the number of queries was 48388 and the execution time of the
attack was 51m13s (less than an hour).

\bheading{Breaking mbedTLS-SGX.}
We also succeeded in carrying out the CBC padding oracle attacks against an
open-source SGX implementation of mbedTLS, mbedTLS-SGX~\cite{mbedtlssgx}, which
can be loaded natively in enclaves. Guided by \frameName, we chose two pages
containing the functions \texttt{sha1\_process\_wrap()} and
\texttt{mbedtls\_sha1\_process()} to monitor.  The template sequence is
``0101...10'' (15 zeros and 14 ones).  Incorrect-padding traces are ``0101...010''
(14 zeros and 13 ones) in all cases.  In our experiment, the attack took
29 minutes and 29 seconds with 25,717 queries to complete the decrypting of one random AES block.

\subsection{Bleichenbacher Attacks}
\label{subsec:bleichenbacher}

\bheading{The oracle.}
Our attack target was the implementation of TLS v1.2 in OpenSSL 1.0.2j (latest
as of February 2017). \frameName identified a vulnerability in the
implementation: the control flows involving \texttt{ERR\_put\_error()}
and \texttt{RSA\_padding\_check\_PKCS1\_type\_2()} may leak sensitive information
regarding the correctness of the formatting. We label the two memory pages that
contain the two functions, respectively, as page 0 and page 1.  The template
sequence is ``1010''. Therefore, if the page sequence in the PTR matches the
template, the \texttt{Oracle()} returns true. Otherwise, in which case the
sequence in the PTR is typically ``10101010'', the oracle returns false.

\bheading{Detailed implementation.}
We use Graphene-SGX to run \textit{unmodified} OpenSSL inside SGX enclaves.
Unlike GnuTLS, OpenSSL does not have enclave-illegal instructions and can be
loaded and ran directly by an SSL/TLS server as \texttt{sgx.trusted\_files} in
the enclave with Graphene.  To complete the attack, we extended the open-source
tool, TLS-Attacker~\cite{Somorovsky:2016:tlsattacker}, and implemented an add-on
module.  We chose TLS-Attacker because it enables us to easily replace the
\texttt{ClientKeyExchange} message with any message we would like the oracle to
test. We did not implement any additional kernel modules besides POAPI, as the
desired computation in kernel space is rather inefficient.  All the attack steps
were accomplished in the userspace with the support of POAPI for querying the
oracle.

With an intercepted \texttt{ClientKeyExchange} message, the attacker
conducts the Bleichenbacher attack to decrypt it and extract the
\texttt{PreMasterSecret}. To do so, the attacker initializes the
man-in-the-kernel attacks through the POAPI module and provides the server
process name, the virtual addresses of the two target pages, and the template
sequence ``1010''. Then the attacker establishes a series of queries: Before
sending each query, he initiates a new TLS handshake with the victim server.
Right before sending the crafted \texttt{ClientKeyExchange} message, the
attacker calls \texttt{Reset()} to POAPI to reset the PTR. Then the crafted
message is sent and the attacker waits until receiving the Alert message from
the server. Then the attacker calls \texttt{Oracle()} to query the oracle,
depending on the return value, the next ciphertext is calculated.  This process
continues until the plaintext of the \texttt{ClientKeyExchange} message is recovered.

\bheading{Evaluation.}
The numbers of queries that are needed to break the \texttt{ClientKeyExchange} message
encrypted with a 1024-bit RSA key, a 2048-bit RSA key and a 4096-bit RSA key are
shown in \tabref{tab:results}. It can be seen that the numbers of queries for
breaking \texttt{ClientKeyExchange} encrypted with 1024-bit key and 2048-bit key
are similar, this is because the oracle strength is not linear in the size of
the keys. Breaking the 2048-bit key encrypted \texttt{ClientKeyExchange} takes
roughly half an hour.  Once the \texttt{PreMasterSecret} is known, the attacker
can decrypt all the intercepted \texttt{application data} packets and hijack
the future communication if the session is still alive. We anticipate an
optimization in the attacks will further speed up the process, possibly making
an online SSL/TLS hijacking attack feasible.

\begin{table}[t]
\begin{minipage}{0.8\columnwidth}
\centering
\caption{\small Performance of the Bleichenbacher attacks against OpenSSL with different key size.}
\label{tab:results}
\end{minipage} \\[-10pt]
\centering \small{
\begin{tabular}[t]{c|c|c|c}
&  1024 & 2048 & 4096\\
\Xhline{1pt}
\textbf{Num. of queries} & 19,346 & 19,368 & 57,286 \\
\Xhline{0.5pt}
\textbf{Time to succeed} & 28m20s& 33m24s & 1h31m39s\\
\end{tabular}} \\[4pt]
\end{table}

% discussion 
\section{Countermeasures}
\label{sec:discuss}

%\bheading{Implementing SSL/TLS in enclaves.}
%Implementing SSL/TLS in SGX enclaves needs more attention than it currently
%receives. One major change of the threat models of the SSL/TLS protocol is the
%shift from a man-in-the-middle adversary to a \textit{man-in-the-kernel}
%adversary.  Man-in-the-kernel attacks consider a malicious OS, who is capable of
%not only conducting man-in-the-middle attacks in the traditional sense, but also
%interacting with SSL/TLS implementation through the kernel/userspace interface,
%such as exceptions and interrupts, system calls, resource allocation, \etc.  The
%additional capabilities enabled by the malicious OS will inevitably make the
%design and implementation of the SSL/TLS protocol more challenging.  Our study
%illustrate one aspect of these challenges. But there are other issues to be
%considered as well. For instance, without placing any trust on the OS kernel,
%the SSL/TLS implementations will find it improbable to reliably validate X.509
%certificates by checking their \texttt{Common Name} fields which use host names
%or IP addresses as server identities. This issue can be addressed by
%short-circuiting the certificate validation process with certificate pinning. A
%better understanding of the man-in-the-kernel attack requires more research
%efforts. 

Countermeasures to the demonstrated attacks can be pursued in three different
layers:

\bheading{Preventing \attackNames.}
Although Intel claims side-channel attacks are outside the threat model of
SGX~\cite{enclaveguide}, given the severity of the demonstrated attacks (among
the others~\cite{Xu:2015:controlled, Shinde:2015:PYF, Lee:2016:SGXbranch}), we
believe it is reasonable for Intel to start exploring solutions to these
side-channel attacks, particularly \attackNames. Some academic research studies
have already made some progress towards this direction~\cite{Shinde:2015:PYF,
Costan:2016:sanctum, Shih:2017:tsgx, Chen:2017:dejavu}.  However, all of these
prior work only considers some of the side-channel attack vectors. But effective
solutions require a complete understanding of the attack surfaces. Due to the lack
of systematic knowledge, none of the prior solutions have successfully prevented
\attackNames on all levels (\ie, page-level, cache-level, and branch-level).
We believe a considerable amount of research in this direction is warranted. 

\bheading{Patching \vulNames.}
An alternative solution is to patch the vulnerabilities in the SSL/TLS
implementations. Constant-time cryptography has been regarded the best practice
to address side-channel issues. But as shown in our study, what create the
oracles are not always the cryptographic operations themselves, but sometimes the error handling
and reporting functions (see \appref{app:examples}). Therefore, although some previous work
has attempted to verify the constant-time implementation in
OpenSSL~\cite{Almeida:2016:VCT, Almeida:2016:VSC}, our attacks suggest that the
entire software package needs to be analyzed together rather than individual
algorithms. By contrast, \frameName can be used to dynamically analyze the whole
software program and pinpoint the vulnerabilities that violate the constant-time
programming paradigm. However, we admit that patching the vulnerability is still
a manual work. For instance, as shown in Listing~\ref{lst:openssl:rsa1} and
Listing~\ref{lst:openssl:rsa2}, the oracle can be removed by eliminating the
\texttt{RSAerr()} call in \texttt{RSA\_padding\_check\_PKCS1\_type\_2()} as 
well as that in its caller function, \texttt{RSA\_eay\_private\_decrypt()}, 
since errors will be reported again after the \texttt{PreMasterSecret} is recognized as invalid. 
If different error types are to be reported, different return values 
could be used to tell them apart. After applying the
patches manually, the SSL/TLS libraries can be tested again using \frameName to
identify the remaining vulnerabilities. Future work should emphasize the
automation of program analysis and patching.

%is capable of helping the developers detect \vulNames for some software
%programs (\eg, SSL/TLS libraries) for which the definition of sensitive
%information is clearly defined (\eg, correctness of padding). However,
%generalization of such methods to arbitrary programs still requires future
%investigation.

%compiler-assisted program transformation, such as~\cite{Rane:2015:raccoon,
%Crane:2015:diversity}, may have the potential to help patch the vulnerabilities
%automatically, but effectiveness in the SGX enclave settings needs to be
%evaluated and the performance overhead must be lowered.

\bheading{Avoiding using the vulnerable ciphersuites.}
The root cause of the Bleichenbacher attacks is the use of the RSA algorithm for
key exchanges, which gives the man-in-the-middle adversary opportunities to
query oracles and decrypt the key materials. It has been shown that both PKCS\#1
v1.5 and PKCS\#1 v2 (\ie, RSAES-OAEP)~\cite{manger2001chosen} are vulnerable to
such attacks. Therefore, to completely mitigate Bleichenbacher attacks,
RSA-based key exchange must be prohibited and the use of DH key exchanges must
be enforced.  To mitigate CBC padding oracle attacks, it is recommended to
replace \texttt{MAC-Then-Encrypt} with the Authenticated Encryption with
Associated Data (AEAD) mode, \eg, AES-GCM.  In the draft version of TLS
v1.3~\cite{TLS13}, it is recommended to use ciphersuites that employ DH for key
exchanges and AEAD modes for symmetric encryption. However, because most SSL/TLS
implementations need to be backward compatible, it may take years before
RSA-based key exchanges and CBC mode symmetric encryptions completely phase out.
Given the severity of the demonstrated attacks in this paper, we recommend
enforcing the use of secure ciphersuites for enclave programs, but it also means
that any entities communicating with the secure enclaves must design special
security policies to disallow the use of these vulnerable ciphersuites also.

%Our study is limited in a sense that we are not able to automatically detect
%unknown types of oracle attacks. All tests are constructed manually by learning
%from prior work. However, we note that automatic detection of such
%vulnerabilities is an open program that none of the prior studies can
%successfully address. TLS-attacker~\cite{Somorovsky:2016:tlsattacker} is also
%limited in the same sense. 

%section related work
\section{Related Work}
\label{sec:related}

%In this section, we summarize related work to our study. We first describe prior
%demonstration of CBC padding oracle attacks and Bleichenbacher attacks. Then we
%introduce several applications of Intel SGX, as well as side-channel attacks and
%defenses mechanisms on SGX. At last, we describe previous studies that focus on
%verification SSL/TLS constant-time execution. 

\subsection{SSL/TLS Oracle Attacks}

\bheading{CBC padding oracle attacks.}
The first discussion of the CBC padding oracle attacks was by Vaudenay in this
seminal paper~\cite{vaudenay2002security}. It was shown that plaintext recovery
is possible without decryption keys if an oracle that differentiates correct
padding from incorrect padding is available to the adversary. Implication on TLS
v1.0 and SSL v3.0 was discussed in the paper, which suggested TLS v1.0 was
potentially exploitable by such attacks because the padding error message is
observable as a reply, but SSL v3.0, due to its unified error message, is more
challenging to exploit. The Vaudenay attack has been mitigated thereafter by
eliminating the error-message oracles. 

The work of Vaudenay was followed up by several studies. Canvel
\etal~\cite{Canvel:2003:PIS} exploited the timing differences in the SSL/TLS
message decryption process as a padding oracle. The timing differences are
caused by the absence of MAC checks when the format of the padding is incorrect. The
countermeasure implemented in popular SSL/TLS libraries, \eg, OpenSSL, is to
compute MAC regardless of the padding correctness.  However, almost 10 years
later, this defense mechanism was circumvented by more sophisticated timing
analysis attacks. AlFardan \etal~\cite{Fardan:2013:lucky} describe a new
padding oracle attack, dubbed Lucky Thirteen Attack, in which the adversary is
able to distinguish the latency of the returned SSL error message
when the number of hash function calls is different. The nuance can serve as
a padding oracle because the correctness of the padding also dictates the number
of hash function calls.  Although the Lucky Thirteen Attack was soon patched
by adding dummy hash operations to enforce constant-time execution, new
vulnerabilities were later discovered by Albrecht
\etal~\cite{albrecht2016lucky}, who proposed new variants of Lucky Thirteen
attacks against the SSL/TLS implementation in Amazon's s2n.  M{\"o}ller
\etal~\cite{Moller:2014:poodle} performed a downgrade attack against SSL/TLS, by
forcing the use of SSL v3.0 during the negotiation of cipher suite in the SSL
handshake protocol. This attack is also called POODLE attacks. SSL v3.0 is
inherently vulnerable to padding oracle attack because its the MAC does not
protect the padding and also padding is nondeterministic except for the last
byte. Due to POODLE attacks, SSL v3.0 has been deprecated.
 
Closer to our study is Irazoqui \etal~\cite{irazoqui2015lucky}, who
demonstrated padding oracle attacks enabled by \flushreload cache side channels.
Our work goes beyond their study in two dimensions: first, we systematically model
various types of control-flow inference attacks under the scenarios of secure
enclaves, instead of considering only cache side-channel attacks.  Second, they
only manually studied the source code of SSL/TLS implementations. In contrast, we
proposed a differential analysis framework to detect vulnerabilities in a wide
range of SSL/TLS implementations, enabling examination of future 
implementations in an automated and black-box fashion. 

\bheading{Bleichenbacher attacks.}
Oracle attacks due to format errors in asymmetric encryption, \ie, RSA
algorithms, can date back to 1998 when Bleichenbacher
\cite{bleichenbacher1998chosen} brought forward the first attack against
PKCS\#1. These attacks rely on oracles of correctly formatted plaintext message
conforming to PKCS\#1 v1.5 standard (\ie, plaintext message must start with
0x0002). PKCS\#1 v2 introduced RSAES-OAEP which employs Optimal Asymmetric
Encryption Padding (OAEP) to mitigate this original Bleichenbacher attack.  A
few years later, it was discovered by Manger~\cite{manger2001chosen} that in
RSAES-OAEP, the length limitation of the plaintext to be encrypted renders its
first byte to be 0x00; failure of conformant to this standard will produce an
error message, which can serve as a new Bleichenbacher oracle. Two years later,
a new, so-called bad-version oracle, was discovered by Klima
\etal~\cite{klima2003attacking} by checking error message regarding incorrect
SSL/TLS version number in the formatted message. The efficiency of the original
Bleichenbacher attack was improved by Bardou~\etal~\cite{bardou2012efficient},
which was also the basis of our attacks. The solution to these attacks was to
unify the error messages so that the adversary is not able to distinguish this
particular format error. To achieve this, TLS v1.0, v1.1, and v1.2 specification
all prescribe that a random number is generated and used as the
\texttt{PreMasterSecret} to enforce approximately equal processing time for both
compliant and non-compliant \texttt{ClientKeyExchange} messages. 

In 2014, Meyer \etal~\cite{meyer2014revisiting} found that SSL/TLS was
vulnerable to timing-based Bleichenbacher attacks. Their attack was enabled by a
timing oracle due to the extra time used to generate pseudo-random numbers when
messages were non-compliant. Most recently, Aviram \cite{aviramdrown} managed to
leverage Bleichenbacher attacks to break TLS 1.2, if the private key is shared
with an SSL/TLS server that supports the legacy SSL v2.0 protocol, which is still
vulnerable to simple Bleichenbacher attacks. A large number of servers were
vulnerable to this so-called DROWN attack. In the non-TLS setting,
Bleichenbacher attacks have been employed to break XML
encryption~\cite{Jager:2012:BAS, Zhang:2014:CSA}. 

Our work suggests that under the scenario of secure enclaves, even the latest
SSL/TLS implementations are vulnerable to Bleichenbacher attacks because the
oracles due to \vulNames are difficult to conceal even in shielded enclaves.

\subsection{Intel SGX: Applications and Attacks}

Intel SGX is a revolutionary technology for applications that require shielded
execution---execution that is isolated from interference or inspection by any
other software components including the privileged system software.  It also
offers remote attestation and sealed storage primitives for trusted
computations. Applications that utilize these new SGX features have been
proposed in previous studies~\cite{Anati:2013:sgxseal, McKeen:2013:sgxisolate,
Hoekstra:2013:sgxsolution, Baumann:2015:haven, Schuster:2015:vc3,
Zhang:2016:towncrier, Tramer:2016:SGP}. Others have been working on facilitating
the development and security protection of SGX
enclaves~\cite{Shinde:2017:Panoply, Seo:2017:sgxshield, Matetic:2017:ROTE}.

Side-channel attacks against SGX enclaves have been described in a few studies.
For example, Xu \etal~\cite{Xu:2015:controlled} and Shinde
\etal~\cite{Shinde:2015:PYF} explored leakage of page-level memory access
pattern due to induced page-fault traces. Lee \etal~\cite{Lee:2016:SGXbranch}
explored processor Branch Target Buffers (BTB) to exploit \vulNames in secure
enclaves. Besides these new attacks in SGX contexts, existing cache attacks are
also applicable against SGX enclaves~\cite{Schwarz:2017:MGE, Brasser:2017:SGE},
since SGX provides no additional protection against such attacks. Defenses
against these attacks are implemented on the hardware
level~\cite{Costan:2016:sanctum} or as compiler
extensions~\cite{Shinde:2015:PYF, Chen:2017:dejavu, Shih:2017:tsgx}.  These
defenses primarily work on page-fault attacks~\cite{Shinde:2015:PYF,
Costan:2016:sanctum, Chen:2017:dejavu, Shih:2017:tsgx} or interrupt-based
side-channel attacks~\cite{Chen:2017:dejavu, Shih:2017:tsgx}. Therefore, they
only remove a portion of the entire attack surface. Completely eliminating
\attackNames that we model in \secref{sec:threat} is extremely challenging.
Therefore, our study of SSL/TLS implementations' \vulNames is not completely
addressed by any of these specific defense techniques.

\subsection{Security Analysis of TLS Implementations}

There has been work on verifying constant-time implementation for SSL/TLS
libraries~\cite{Almeida:2016:VCT, Almeida:2016:VSC}. However, our findings
suggest that control-flow leakages still exist even when constant-time mechanisms
are employed, especially when the constant-time implementation is enforced by
making dummy function calls which may include control flows that depend on the
error types. We also find leakage occurs when the internal error
logging and reporting functions reveal the reasons for the errors. 

%such as in OpenSSL. Several common 
%problems include extra function calls to conserve constant-time and the extra error 
%logging. Examples of such vulnerabilities can be found in \ref{app:examples}. The 
%strength of our tool enables us to automatically find unknown side-channel vulnerabilities 
%in the view of control flow without even reading any code. The trivial ones not 
%obversant in the timing channel would be exposed.

Differential analysis has been applied to examine the implementation of
certificate validation in TLS libraries~\cite{Brubaker:2014:UFA, Chen:2015:GDT}.
Others focus on determining whether a TLS implementation correctly follows the
TLS protocol~\cite{Somorovsky:2016:tlsattacker, de2015protocol,
beurdouche2015messy}. Our work is different from them both in the design goals and the
methodologies.

%But they all 
%focuses on determining whether an TLS implementation is correctly built following the 
%TLS protocol. TLS-Attacker~\cite{Somorovsky:2016:tlsattacker} is used in our work as part 
%of \frameName. It provides a feature called modifiable variables to allow modifications 
%to any TLS packets to be sent. Ruiter~\cite{de2015protocol} used state machine in a fuzzing 
%%fashion to find potential flaws in program logic from both server-side and client- side. 
%FLEXTLS~\cite{beurdouche2015messy} is a tool to evaluate TLS state machines with custom 
%variable values and protocol message flows. Two other tools~\cite{tlsfuzzer, scappy} 
%can also be used to implement stateless protocol executions and variable customizations.

%Particularly, Frankencert~\cite{Brubaker:2014:UFA} is an automated method for finding 
%certificate validation vulnerabilities by synthesizing randomly mutated certificates. 
%Mucert introduced in \cite{Chen:2015:GDT} is a guided version of Frankencert. Adopting 
%Markov Chain Monte Carlo (MCMC), it has better efficiency and wider coverage than 
%Frankencert. 

\section{Conclusion}
\label{sec:conclude}

In this paper, we studied oracle attacks against SSL/TLS implementations in SGX.
These attacks are enabled by \vulNames in SSL/TLS libraries and are exploitable
by branch-level, cacheline-level, and page-level \attackNames. Our implementation
of man-in-the-kernel attacks empirical demonstrated that the resulting oracle
attacks are highly efficient. We also designed a differential analysis framework
to help detect these vulnerabilities automatically. We show that all the
open-source SSL/TLS libraries we examined are exploitable, thus raising the
questions of secure development and deployment of SSL/TLS in SGX enclaves.

\section*{Acknowledgement}
We are grateful to the anonymous reviewers for their constructive comments. This work was supported in part by NSF 1566444.

\bibliographystyle{ACM-Reference-Format}
\bibliography{paper}

\appendix

\section{Examples of \vulNames}
\label{app:examples}

\subsection{Padding Oracles in mbedTLS(-SGX)}
In mbedTLS v2.4.1 and mbedTLS-SGX, the decryption oracle can be constructed by
monitoring the function \texttt{sha1\_process\_wrap()} and \texttt{mbedtls\_sha1\_process()}.
Particularly, as shown in Listing~\ref{lst:mbedtls:cbc}, the function
\texttt{ssl\_decrypt\_buf()} calls \texttt{mbedtls\_md\_process()}, which is a
wrapper function that calls both \texttt{sha1\_process\_wrap()} and
\texttt{mbedtls\_sha1\_process()}, to conceal the timing difference caused by removing the 
paddings before calculating the MAC. However, the number
of times \texttt{mbedtls\_md\_process()} is called depends on the value of
\texttt{extra\_run}, which is calculated from the length of the padding,
\texttt{padlen}. Particularly, when the padding is incorrect, \texttt{padlen}
will be 0, and \texttt{mbedtls\_md\_process()} is called only once. Therefore, the
number of calls to \texttt{sha1\_process\_wrap()} and
\texttt{mbedtls\_sha1\_process()}, which are located on different pages, has
been exploited as the oracle in our demonstrated attacks.
We note that the padding oracle is created due to the improper constant-time
implementation of defenses to existing attacks.

\begin{figure}[t]
\centering
\begin{minipage}{.96\columnwidth}
\begin{lstlisting}[caption={Snippet of \texttt{ssl\_decrypt\_buf()}},
label={lst:mbedtls:cbc}]
    ...

    padlen &= correct * 0x1FF;
    ...

    size_t j, extra_run = 0;
    extra_run = ( 13 + ssl->in_msglen + padlen + 8 ) / 64 
	       - ( 13 + ssl->in_msglen + 8 ) / 64;

    extra_run &= correct * 0xFF;

    mbedtls_md_hmac_update( &ssl->transform_in->md_ctx_dec, ssl->in_ctr, 8 );
    mbedtls_md_hmac_update( &ssl->transform_in->md_ctx_dec, ssl->in_hdr, 3 );
    mbedtls_md_hmac_update( &ssl->transform_in->md_ctx_dec, ssl->in_len, 2 );
    mbedtls_md_hmac_update( &ssl->transform_in->md_ctx_dec, ssl->in_msg,
                     ssl->in_msglen );
    mbedtls_md_hmac_finish( &ssl->transform_in->md_ctx_dec,
                     ssl->in_msg + ssl->in_msglen );
    /* Call mbedtls_md_process at least once due to cache attacks */
    for( j = 0; j < extra_run + 1; j++ )
        mbedtls_md_process( &ssl->transform_in->md_ctx_dec, ssl->in_msg );
    ...
\end{lstlisting}
\end{minipage}
\vspace{-0pt}
\end{figure}

\subsection{Bleichenbacher Attack Oracles in OpenSSL}

\begin{figure}[t]
\centering
\begin{minipage}{.96\columnwidth}
\begin{lstlisting}[caption={\texttt{RSA\_eay\_private\_decrypt()}},
label={lst:openssl:rsa1}]
	...

	switch (padding) {
		case RSA_PKCS1_PADDING:
	        r = RSA_padding_check_PKCS1_type_2(to, num, buf, j, num);
	        break;
	    	...
	}
	if (r < 0)
	        RSAerr(RSA_F_RSA_EAY_PRIVATE_DECRYPT,
			RSA_R_PADDING_CHECK_FAILED);
	...
\end{lstlisting}
\end{minipage}
\vspace{-0pt}
\end{figure}

\begin{figure}[t]
\centering
\begin{minipage}{.96\columnwidth}
\begin{lstlisting}[caption={\texttt{RSA\_padding\_check\_PKCS1\_type\_2()}},
label={lst:openssl:rsa2}]
	...

	if (tlen < 0 || flen < 0)
	        return -1;
	if (flen > num)
	        goto err;
	if (num < 11)
	        goto err;
	...

	good = constant_time_is_zero(em[0]);
	good &= constant_time_eq(em[1], 2);
	...

	good &= constant_time_ge((unsigned int)(zero_index), 2 + 8);
	...

	good &= constant_time_ge((unsigned int)(tlen), (unsigned int)(mlen));
	if (!good) {
	        mlen = -1;
	        goto err;
	}

	err:
	if (em != NULL)
	        OPENSSL_free(em);
	if (mlen == -1)
	        RSAerr(RSA_F_RSA_PADDING_CHECK_PKCS1_TYPE_2,
	               RSA_R_PKCS_DECODING_ERROR);
	return mlen;
\end{lstlisting}
\end{minipage}
\vspace{-0pt}
\end{figure} 

The oracle in OpenSSL 1.0.2j is created by function \texttt{RSAerr()}. As shown
in Listing~\ref{lst:openssl:rsa2}, in
\texttt{RSA\_padding\_check\_PKCS1\_type\_2()}, when any error is detected
during the PKCS decoding procedure, \texttt{mlen} will be set to -1. Thus
\texttt{RSAerr()} will be called to report the error before the function
returns. After returning to the caller function \texttt{RSA\_eay\_private\_decrypt()}
(shown in Listing~\ref{lst:openssl:rsa1}), \texttt{RSAerr()} is called one more
time. 
%However, for both conformant and non-conformant formats,
%an error of wrong PMS will be reported later after further computation with PMS fails. 
These two calls to the \texttt{RSAerr()} reveals a non-PKCS-conformant
formatting, which can be exploited as an oracle for Bleichenbacher attacks.
The vulnerabilities in OpenSSL is not because of a failed constant-time
implementation, but the redundant error reporting and logging mechanisms. One
possible suggestion is to avoid repeated error reporting that are due to
different reasons.

%In fact, \texttt{RSAerr()} will be called again later when the checking
%of PMS fails, for both conformant and non-conformant PKCS formats. 

\subsection{Bleichenbacher Oracles in GnuTLS}

\begin{figure}[t]
\centering
\begin{minipage}{.96\columnwidth}
\begin{lstlisting}[caption={Snippet of \texttt{proc\_rsa\_client\_kx()}},
label={lst:gnutls:rsa}]
	...

	if (ret < 0 || plaintext.size != GNUTLS_MASTER_SIZE) {
		_gnutls_debug_log("auth_rsa: Possible PKCS #1 format attack\n");
		use_rnd_key = 1;
	} else {
		if (_gnutls_get_adv_version_major(session) !=
	plaintext.data[0] || 
	(session->internals.priorities.allow_wrong_pms == 0
	&& _gnutls_get_adv_version_minor(session) !=
		plaintext.data[1])) {
			_gnutls_debug_log("auth_rsa: Possible PKCS #1 version check format attack\n");
		}
	}
	...
\end{lstlisting}
\end{minipage}
\vspace{-0pt}
\end{figure} 

Similar to OpenSSL, the RSA decryption oracle in GnuTLS is also due to error
logging and reporting. As shown in Listing~\ref{lst:gnutls:rsa},
the function \texttt{\_gnutls\_debug\_log()} is called for either an incorrect PKCS format
or incorrect version numbers. Although GnuTLS applies the countermeasure
against Bleichenbacher attack by using random \texttt{PreMasterSecrets}, the
logging functions expose the error messages to the adversary with the
capability of conducting \attackNames. 

%\bheading{Summary:} The oracle is created not because of improper implementation
%of constant-time cryptography, but error logging and reporting functions. 

\end{document}